\documentclass[lettersize,journal]{IEEEtran}
\usepackage{amsmath,amsfonts}
\usepackage{algorithmic}
\usepackage{algorithm}
\usepackage{array}
\usepackage[caption=false,font=normalsize,labelfont=sf,textfont=sf]{subfig}
\usepackage{textcomp}
\usepackage{stfloats}
\usepackage{url}
\usepackage{verbatim}
\usepackage{graphicx}
\usepackage{cite}
\usepackage{pifont}  %
\usepackage{xcolor}  %
\definecolor{bedroom_color}{RGB}{251,154,153}
\definecolor{kitchen_color}{RGB}{178,223,128}
\newcommand{\cmark}{{\color{kitchen_color}\ding{51}}}  %
\newcommand{\xmark}{{\color{bedroom_color}\ding{55}}}  %

\begin{document}

\title{Eliminating Rasterization: Direct Vector \\ Floor Plan Generation with DiffPlanner}

\author{Shidong~Wang and Renato~Pajarola,~\IEEEmembership{Senior Member,~IEEE}
\thanks{S.~Wang and R.~Pajarola are with the Department of Informatics, University of Zurich, Switzerland. E-mails: \{shwang,~pajarola\}@ifi.uzh.ch.}
\thanks{Received 30 July 2024; revised 12 March 2025; accepted 5 April 2025.}}

\markboth{IEEE Transactions on Visualization and Computer Graphics}%
{Wang \MakeLowercase{\textit{et al.}}: Eliminating Rasterization: Direct Vector Floor Plan Generation with DiffPlanner}


\maketitle
\begin{abstract}

The boundary-constrained floor plan generation problem aims to generate the topological and geometric properties of a set of rooms within a given boundary.
Recently, learning-based methods have made significant progress in generating realistic floor plans.
However, these methods involve a workflow of converting vector data into raster images, using image-based generative models, and then converting the results back into vector data.
This process is complex and redundant, often resulting in information loss.
Raster images, unlike vector data, cannot scale without losing detail and precision.
To address these issues, we propose a novel deep learning framework called DiffPlanner for boundary-constrained floor plan generation, which operates entirely in vector space.
Our framework is a Transformer-based conditional diffusion model that integrates an alignment mechanism in training, aligning the optimization trajectory of the model with the iterative design processes of designers.
This enables our model to handle complex vector data, better fit the distribution of the predicted targets, accomplish the challenging task of floor plan layout design, and achieve user-controllable generation.
We conduct quantitative comparisons, qualitative evaluations, ablation experiments, and perceptual studies to evaluate our method.
Extensive experiments demonstrate that DiffPlanner surpasses existing state-of-the-art methods in generating floor plans and bubble diagrams in the creative stages, offering more controllability to users and producing higher-quality results that closely match the ground truths.

\end{abstract}

\section{Introduction} \label{sec:introduction}

\IEEEPARstart{D}{ata-driven} boundary-constrained floor plan generation, which attempts to generate the geometric and topological properties of a set of rooms given the interior of a boundary, has recently attracted widespread interest within the computer graphics, vision, and architecture communities~\cite{Wu-19, Chaillou-20, Hu-20, Wang-21, He-22, Sun-22}.
Floor plan generation represents a typical coarse-to-fine layout design process.
The input to this task is the exterior boundary of a building, and the output should contain the number, categories, locations, adjacencies, and partitioning of rooms inside the boundary.
Also, the output is expected to be in the format of vector data, which is critical for practical use by architects and designers, and which is also the reason why a large number of researchers have investigated the vectorization of raster floor plans~\cite{liu-17, Hu-24}.

\begin{figure*}[t]
    \centering
    \includegraphics[width=\linewidth]{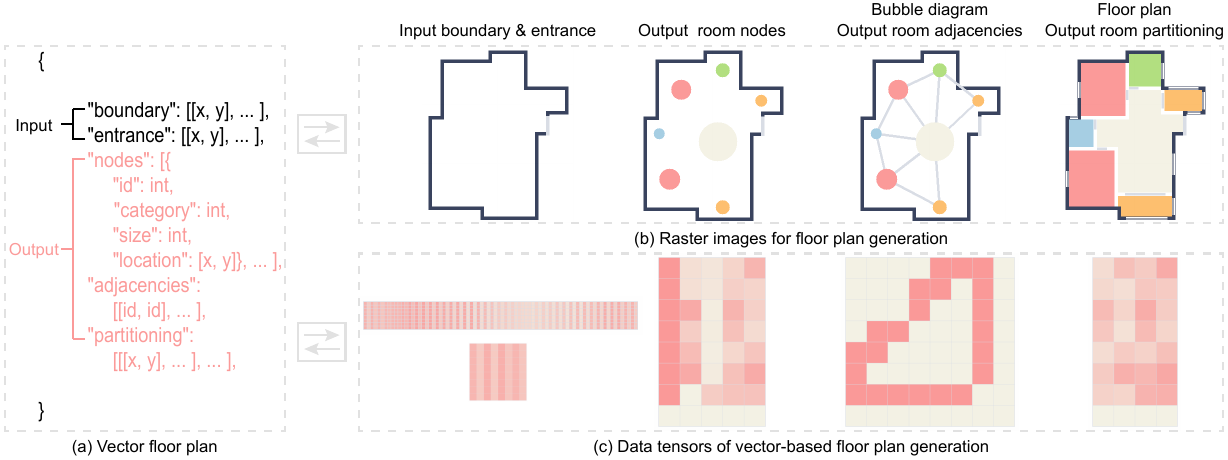}
    \caption{Most previous methods with input and output vector data (a), internally rasterize the input into a series of images representing the layout and then use image-based networks for the prediction process followed by vectorization of the output (b). In contrast, our DiffPlanner performs the prediction process directly based on data tensors representing the vector space (c).}
    \label{fig:teaser}
\end{figure*}

As illustrated in Figure~\ref{fig:teaser}(b), most of the existing previous methods have adopted a raster-to-raster generation pipeline~\cite{Wu-19, Chaillou-20, Wang-21, Sun-22}, i.e., first rasterizing the boundary information from a vector dataset, and then using an image-based generative model (e.g., CNN or GAN) to directly generate a raster image that contains the geometric and topological information of the room(s), and eventually extracting the vector information from this raster output.
Recently, some approaches have used a hybrid strategy, i.e., using graph neural networks (GNN) or Transformer, to directly predict vector information about the layout, such as the coordinates of a bounding box representing a room~\cite{Hu-20, He-22}, or a bubble diagram containing the locations of nodes and their adjacencies~\cite{Sun-23}.
However, they still depend on image-based generative models, e.g., to perform feature extraction on the input raster boundaries and the ground truth raster images~\cite{Hu-20, He-22, Sun-23}, or to generate a raster layout image to provide additional information for floor plan prediction~\cite{Hu-20, He-22}. 

Existing methods could generate some results comparable to those of professional architects, thanks to the carefully designed models.
However, their workflow of converting vector information into images, then performing the generation process via an image-based generative model, and finally converting the raster results into vector data is unnecessarily redundant and overly complex.
Moreover, when converting vector data, like floor plans, to raster images, information loss often occurs~\cite{Congalton-97, Huo-22}.
Vector data, which is defined by precise geometry and topology, can scale infinitely without distortion.
In contrast, raster images are made of fixed-resolution pixels, which can result in lost details and precision, and are scale-dependent.
Common issues include blurred edges, altered angles, broken lines, and oversimplified complex structures during the rasterization process.

Based on this, we considered whether a model could be developed that learns directly from vector data and outputs layouts in the same format.
This proves to be a challenging task.
For the input, learning directly from vector data is difficult as it is not as standardized as learning from raster images.
To illustrate this, when faced with raw vector data (Figure~\ref{fig:teaser}(a)), humans often have only a superficial understanding, but a raster image (Figure~\ref{fig:teaser}(b)) allows us to quickly grasp basic topological and geometric information.
For the output, we expect the model not only to adapt to the early stages of design, offering rough designs that include geometric and topological information (e.g., bubble diagrams), but also to produce more detailed layout results (e.g., accurate floor plans).
The output needs to be significantly more complex than the input boundaries, meaning that the model must be capable of learning the distribution from more complex vector data and ultimately producing detailed vector layouts from simple inputs (e.g., boundaries).
Regarding the generation process, we anticipate that the model will be fully automatic, capable of autonomously generating the complete floor plan layouts without any information beyond the boundaries.
However, the process should be controllable on a high level, allowing users to input simple conditions such as the number of rooms and their categories.
Additionally, it should be adjustable, permitting iterative user interventions at each step of the design.

Based on this, we propose DiffPlanner, a novel deep learning framework for boundary-constrained floor plan generation.
Our framework leverages recent popular diffusion models (DM)~\cite{Ho-20}, which have shown remarkable success in various generative tasks, to handle the non-trivial challenge of floor plan layout design.
It incorporates a Transformer~\cite{Vaswani-17} as the backbone of the model, enabling the handling of complex vector layout data.
Additionally, it integrates a conditioning mechanism that allows for controlled floor plan generation under specific conditions.
We have designed an alignment mechanism during the training phase of the model, which enables the model to extract information from the iterative design processes of designers, thereby better fitting the distribution of predicted targets.

Moreover, our approach also supports generating bubble diagrams and floor plans without predefined boundaries.
Previous methods that do not support predefined boundaries have been significantly limited in terms of user controllability -- they either fully automate the generation process without allowing user intervention~\cite{Sun-23} or only support bubble diagram-driven floor plan generation~\cite{Nauata-20, Nauata-21, Shabani-23}, neglecting the generation of bubble diagrams representing the early design stages.
Compared to previous methods, our new approach produces higher-quality results that are closer to the ground truths and supports various forms of user interaction.

Overall, our contributions are as follows:

\begin{enumerate}
\item 
We frame the task of boundary-constrained floor plan generation as a generative task fully in vector space, for the first time, and propose a generation framework based on diffusion models, which directly predicts vector floor plans from vector boundaries.
Previous methods depended (internally) on raster space, requiring the conversion of vector information into a set of raster images, followed by prediction processes using image-based generative CNN or GAN models.

\item
We have also designed an alignment mechanism to better align the optimization trajectory of the model with the common iterative manual design processes during the training stage.
	
\item 
Our framework can address a coarse-to-fine floor plan layout design problem, even for boundary-unconstrained floor plan generation.
It produces bubble diagrams that represent geometric and topological information of design elements in the early stages, and detailed floor plans in the later stages.
Our approach supports various levels of user interaction, including fully automatic, coarsely controllable, and finely controllable options.
Compared to previous methods, our approach can generate more layout information and provide users with richer interactions.
\end{enumerate}

\section{Related work} \label{sec:related_work}

\subsection{Bubble Diagram Generation} \label{subsec:related_work_bubble}

Bubble diagrams establish a connection between the intent and solution of designer, and they are widely used in data-driven layout planning tasks such as architectural design and interior scene synthesis~\cite{Nauata-20, Hu-20, Nauata-21, Shabani-23}.
In previous work, they often serve as an input condition representing user intent, or as an intermediate representation between input conditions and final layout prediction.
Almost no work focuses on data-driven bubble diagram generation.
Recently, Sun et al.~\cite{Sun-23} proposed a data-driven approach (BubbleFormer) for generating bubble diagrams to better drive downstream layout planning tasks.
BubbleFormer requires rasterizing the input boundaries and the target bubble diagrams into images, then using a CNN to extract features and subsequently performing the prediction process.
Finally, it needs to employ the Hungarian matching algorithm~\cite{Kuhn-55} to obtain unique matches between the generated nodes and edges.

While BubbleFormer can automatically generate diverse results for users to browse and select their preferred designs, the entire generation process does not allow for user interaction.
But interactivity is very useful for an AI tool focused on layout planning~\cite{Shneiderman-20}.
Unlike BubbleFormer, our method supports various forms of user interaction, as shown in Table~\ref{tab:existing_methods}.
It allows users to simply input instructions, such as the number and categories of nodes, and also supports more detailed interactions, such as iterative design of nodes and adjacencies.
Additionally, our method does not require redundant rasterization operations.
All predictions are made directly in vector space, and no post-processing is needed.

\begin{table*}[ht]
    \centering
    \caption{Compared to existing methods for boundary-constrained bubble diagram and floor plan generation, our approach is based on state-of-the-art generative models, specifically diffusion models (DM), and performs predictions directly in vector space without any intermediate rasterization operations. Our method supports the most extensive range of user interactions, from fully automatic generation, where a vector layout can be obtained only from the input boundary, to coarsely controllable prediction, where users can provide simple instructions to specify room nodes ($R_{node}$), room adjacencies ($R_{adja}$), and room partitioning ($R_{part}$), to finely controllable design, where users can iteratively adjust each step of the design process.}
    \label{tab:existing_methods}
    \begin{tabular}{|l|c|cccccccc|}
        \hline
        Bubble Diagram & Network & w/o & Fully & \multicolumn{3}{c}{Coarsely Controllable} & \multicolumn{3}{c|}{Finely Controllable} \\
        Generation & Architecture & Rasterization & Automatic & $R_{node}$ & \multicolumn{2}{c}{$R_{adja}$} & $R_{node}$ & \multicolumn{2}{c|}{$R_{adja}$} \\
        \hline
        BubbleFormer~\cite{Sun-23} & CNN~\&~Transformer & \xmark & \cmark & \xmark &\multicolumn{2}{c}{\xmark} & \xmark &\multicolumn{2}{c|}{\xmark} \\
        DiffPlanner [Our] & DM & \cmark & \cmark & \cmark &\multicolumn{2}{c}{\cmark} & \cmark &\multicolumn{2}{c|}{\cmark} \\
        \hline
        Floor Plan & Network & w/o & Fully & \multicolumn{3}{c}{Coarsely Controllable} & \multicolumn{3}{c|}{Finely Controllable} \\
        Generation & Architecture & Rasterization & Automatic & $R_{node}$ & $R_{adja}$ & $R_{part}$ & $R_{node}$ & $R_{adja}$ & $R_{part}$ \\
        \hline
        ArchiGAN~\cite{Chaillou-20} & GAN & \xmark & \xmark & \xmark & \xmark & \xmark & \xmark & \xmark & \xmark \\
        ActFloor-GAN~\cite{Wang-21} & GAN & \xmark & \cmark & \xmark & \xmark & \xmark & \xmark & \xmark & \xmark \\
        WallPlan~\cite{Sun-22} & CNN & \xmark & \cmark & \xmark & \xmark & \xmark & \xmark & \xmark & \xmark \\
        RPLAN~\cite{Wu-19} & CNN & \xmark & \cmark & \cmark & \xmark & \xmark & \cmark & \xmark & \xmark \\
        Graph2Plan~\cite{Hu-20} & CNN~\&~GNN & \xmark & \cmark & \cmark & \cmark & \xmark & \xmark & \xmark & \xmark \\
        iPLAN~\cite{He-22} & CNN~\&~GAN & \xmark & \cmark & \cmark & \xmark & \cmark & \cmark & \xmark & \cmark \\
        DiffPlanner [Our] & DM & \cmark & \cmark & \cmark & \cmark & \cmark & \cmark & \cmark & \cmark \\
        \hline
    \end{tabular}
\end{table*}

\subsection{Boundary-constrained Floor Plan Generation} \label{subsec:related_work_floorplan_boun}

Floor plan generation has been a prominent topic in the intersection of computer graphics~\cite{Merrell-10, Bao-13, Wu-18} and architecture~\cite{Arvin-02, Michalek-02, Rodrigues-13} for decades, aiming to achieve automated design of architectural floor plans through computational design methods.
The exterior boundary of a building is a crucial condition for floor plan generation~\cite{Wu-19}.
In recent years, more and more researchers have started using learning-based methods to tackle the problem of boundary-constrained floor plan generation~\cite{Wu-19, Chaillou-20, Hu-20, Wang-21, He-22, Sun-22}.

Early researchers directly employed image-based generative models, such as Pix2Pix~\cite{Isola-17}, to predict a rasterized floor plan image from a rasterized boundary image.
However, the generated results were often too noisy to be vectorized~\cite{Chaillou-20}.
Subsequently, some researchers began dividing the floor plan generation process into multiple sub-tasks~\cite{Wu-19, Wang-21, Sun-22}.
Starting with a rasterized boundary, they used image-based models to first predict intermediate raster representations, such as location masks, activity maps, or wall maps, and then predicted rasterized floor plans from these intermediate representations, resulting in more reasonable outputs that could be vectorized.
Next, some researchers attempt to represent floor plans not as raster images but as geometric vectors of the bounding box coordinates~\cite{Hu-20, He-22}.
However, these approaches still required starting from a rasterized boundary and generating a rasterized layout image as auxiliary information for the final layout prediction.

Unlike all previous methods, our approach defines the boundary-constrained floor plan generation problem as a vector-to-vector generation task, directly predicting the vector floor plan from the vector boundary without any rasterization operations, as shown in Table~\ref{tab:existing_methods}.
This enables our model to generate higher-quality and more reasonable floor plans.
Additionally, our model supports more diverse user-controlled generation options, enhancing interactivity compared to previous methods.
This makes our approach more aligned with the user needs and preferences in the design process.

\subsection{Floor Plan Generation without Boundary Constraints} \label{subsec:related_work_floorplan_bubble}

Floor plan generation without boundary constraints is often driven by bubble diagrams.
Bubble diagrams effectively capture the design intent of architects during the initial stages of design.
Therefore, some learning-based approaches have been proposed to enable architectural floor plan generation by mapping bubble diagrams to floor plans~\cite{Nauata-20, Nauata-21, Shabani-23}.
However, these methods typically focus only on the final design stage, converting already finalized design ideas into the final layout, while neglecting the early creative stages, such as the bubble diagram generation phase.

Architects indeed need an automated tool to help convert bubble diagrams, which convey design intent, into final floor plans to reduce their workload.
However, they also need an AI design tool that collaborates with them in the creative stage, providing more inspiration and reducing their workload in conceptual design~\cite{Shneiderman-20}.
Unlike previous methods, our proposed approach collaborates with users from the initial creative stage to the final design stage, first creating bubble diagrams and then generating the final floor plans.
Compared to the state-of-the-art method in bubble diagram-driven floor plan generation, HouseDiffusion~\cite{Shabani-23}, our approach produces higher-quality results that are closer to the ground truths and supports richer user interactions.
\section{Overview} \label{sec:overview}

\begin{figure}[t]
    \centering
    \includegraphics[width=\linewidth]{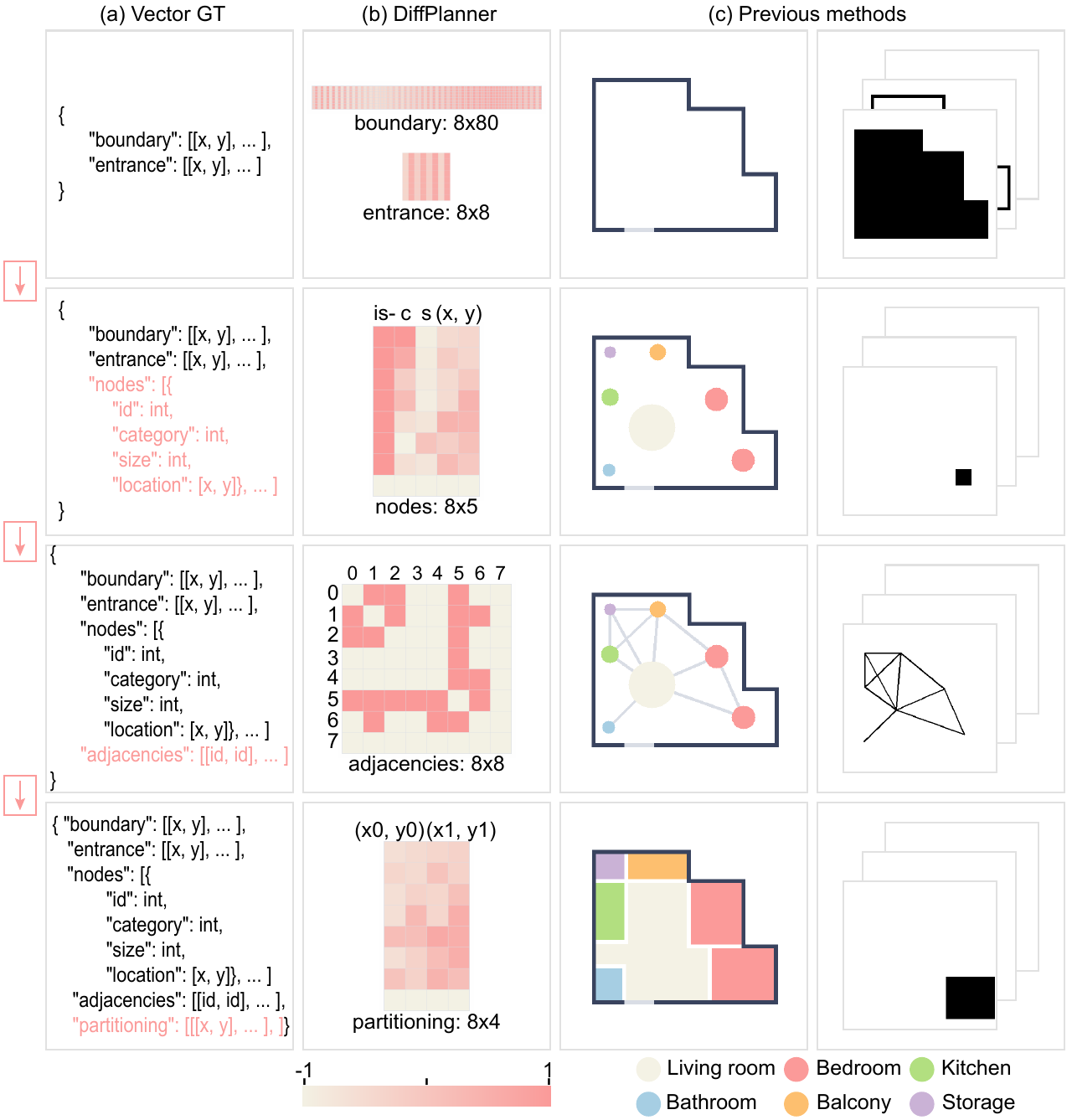}
    \caption{Overview and data representation of DiffPlanner and previous methods. Our DiffPlanner decomposes the floor plan design process into several stages based on established design principles, allowing for human input at various levels of detail. Unlike previous methods that use raster images (c), which require the rasterization of vector data, our approach operates directly on data tensors representing the vector space (b). Starting with input vectors for the room boundary and entrance, our model first predicts room nodes including their number, categories, sizes, and rough locations. It then determines the adjacency matrix between these nodes to form a bubble diagram, representing the initial design phase. Finally, our model predicts the room partitioning to complete the floor plan.}
    \label{fig:overview}
\end{figure}

\emph{Problem.}
The goal of boundary-constrained floor plan generation is to generate a set of geometric and topological properties of rooms within the given boundary.
A floor plan can be considered as a collection of boundary, entrance, and a set of rooms, including information on their categories, sizes, locations, and other details (see Figure~\ref{fig:overview}(a)).
Previous approaches often re-frame the floor plan generation task as an image generation task, which first requires rasterizing the vector floor plans into images representing various layout information (see Figure~\ref{fig:overview}(c)).
Subsequently, a series of image-based networks perform the prediction and generation tasks.
In contrast, we define this task directly as a vector-to-vector generation task (with data tensors as shown in Figure~\ref{fig:overview}(b)).

\emph{Dataset.}
This work utilizes the RPLAN dataset~\cite{Wu-19}, which is a large open-source dataset that contains $80K$ real-world floor plans.
We use the same data preprocessing method as in~\cite{Hu-20} to extract the necessary vector information, such as the room numbers, room categories, room sizes, room locations, room adjacencies, and other relevant details (Figure~\ref{fig:overview}(a)).
We further process this information into a data tensor format that our vector-based model can accept (Figure~\ref{fig:overview}(b)) to obtain the final dataset including $80K$ pairs of data, and randomly split it into $56K$-$12K$-$12K$ for training-validation-test sets.

\emph{Challenge.}
Achieving our goal is non-trivial.
Firstly, it is challenging for the model to directly extract information from vector data and learn layout planning capabilities.
For example, humans find it easier to capture initial layout information from images compared to structured vector data (Figure~\ref{fig:overview}(a) vs. Figure~\ref{fig:overview}(c)).
Secondly, we expect this model to handle diverse layout planning tasks, such as generating bubble diagrams in the early design stages, producing final floor plans in the later stages, and even tackling layout planning tasks without boundary constraints.
Finally, we want this model to deeply interact with users.
It should not only be fully automatic but also allow for coarse user control, where users can provide simple instructions, and fine control, where users can iteratively control the process.
Overall, the model is expected to output complex layouts from simple input information, with the entire generation process being highly controllable by the user and performed entirely in vector space.

\emph{Methodology.}
Similar to existing methods~\cite{He-22, Wu-19}, we also appropriately decompose the floor plan design procedure into several stages based on fundamental design principles and widely adopted practices~\cite{Rengel-16}, allowing for human input at various levels of detail.
Note that, unlike previous methods, our approach performs predictions entirely in vector space, without the need to rasterize the vector information.
As shown in Figure~\ref{fig:overview}(b), starting from a data tensor storing the input vector boundary, that includes the entrance, our model first predicts a set of room nodes including the information about number, categories, sizes, and approximate locations.
Next, it predicts the adjacency matrix between these nodes to then create a bubble diagram, reflecting the initial design stage.
Subsequently, our model further predicts the room partitioning to generate the final floor plan.
We allow the user to interact with the model throughout the entire generation process, such as specifying the number and categories of rooms, iteratively controlling the room locations and their adjacencies, and determining the partitioning of rooms.

\section{Method} \label{sec:method}

In this section, we first provide a brief introduction to the diffusion models (DM) and then explain how our method progressively predicts room nodes, room adjacencies, and room partitioning from the input boundary.

\subsection{Diffusion Models (DM)} \label{subsec:dm}

The diffusion model~\cite{Ho-20} is a generative framework that gradually transforms a simple initial distribution into a complex target distribution through a series of iterative steps.
This model operates by adding noise to the data and then learning to reverse this process, effectively denoising to generate realistic samples.
It is particularly effective for capturing intricate patterns and dependencies in data, making it suitable for tasks that require fine-grained detail and structure.
Specifically, diffusion models (DM) transform Gaussian noise $x_T$ into a data sample $x_0$ through a series of $T$ denoising steps, involving both forward and reverse processes during training.
In the forward process $q(x_t | x_0)$, a data sample $x_0$ is progressively converted into a noisy sample $x_t$ at each time step $t$ by adding Gaussian noise $\epsilon \sim \mathcal{N}(0,\mathbf{I})$:

\begin{equation}
    x_t = \sqrt{\bar \alpha_t}x_0 + \sqrt{(1-\bar \alpha_t)}\epsilon.
    \label{eq:forward}
\end{equation}
where $\alpha_t = (1 - \beta_t)$, $\beta_t$ is the variance schedule controlling the noise level, and $\bar{\alpha}_t = \prod_{s=1}^t \alpha_s$ represents the cumulative data preservation.
The reverse process $p_{\theta}(x_{t-1} | x_{t})$ begins with pure Gaussian noise $x_T \sim \mathcal{N}(0,\mathbf{I})$, and iteratively denoises it step by step until it reaches $x_0$.
During this process, the model takes $x_t$ and estimates $x_{t-1}$ by inferring $\epsilon$~\cite{Ho-20}:

\begin{equation}
    x_{t-1} = \frac{1}{\sqrt{\alpha_t}}(x_t - \frac{1-\alpha_t}{\sqrt{1-\bar \alpha_t}}\epsilon_\theta(x_t, t))+\sigma_t z.
    \label{eq:reverse}
\end{equation}
where ${\epsilon}_{\theta}$ is a function approximator and is optimized by: 

\begin{equation}
    \mathcal{L}_{\text{t-1}} = || {\epsilon}_{\theta} (x_t, t) - {\epsilon} ||^2.
    \label{eq:diff_loss}
\end{equation}

\subsection{Node Generation} \label{subsec:nodediff}

\emph{Problem.}
In this task, we aim to predict vector room nodes including the number, categories, sizes, and locations of rooms from a given vector boundary with entrance.

\emph{Representation.}
The vector boundary and entrance are represented by a set of corner coordinates.
Previous methods~\cite{Wu-19, Chaillou-20, Hu-20, Wang-21, Sun-22, He-22, Sun-23} often rasterize the vector boundary and entrance into a set of binary raster masks and then use an image-based network to extract features (Figure~\ref{fig:overview}(c)).
In contrast, we directly feed the vector boundary and entrance into our model, avoiding the redundant step of rasterizing the boundary and entrance to extract their embeddings.
Note that, while the entrance has a fixed number of four corner points, the number of boundary corner points is variable.
To handle this, we employ a simple corner augmentation strategy for the boundary.
Specifically, we add a corner point at the midpoint of the longest edge of the boundary iteratively until the number of corner points reaches $40$, which is sufficient to represent the exterior boundary of almost all real-world buildings.
Next, we replicate the boundary and entrance eight times, corresponding to the maximum number of rooms in the RPLAN dataset~\cite{Wu-19}.
Finally, we convert the vector boundary and entrance into tensors with dimensions $8 \times 80$ and $8 \times 8$, respectively.

Previous methods typically represent vector room nodes as a multi-channel binary image~\cite{Wu-19, He-22, Sun-23}, using channel indices to indicate categories and masks to represent the sizes and locations of rooms, or as a three-channel color image~\cite{Chaillou-20, Wang-21}, using different colors to denote different categories (Figure~\ref{fig:overview}(c)).
These rasterized results are then used as prediction targets.
Unlike these previous methods, we directly represent the vector room nodes as an $8 \times 5$ tensor in continuous space.
Each row represents information for a single room, and the five columns indicate is-room ($1$ for room, $0$ for padding), category ($1$ to $6$ representing the living room, bedroom, kitchen, bathroom, balcony, and storage), size, $x$ and $y$ coordinates of the location of the room.
Ultimately, we normalize all the data tensors to a range of $[-1, 1]$.

\begin{figure}[t]
    \centering
    \includegraphics[width=\linewidth]{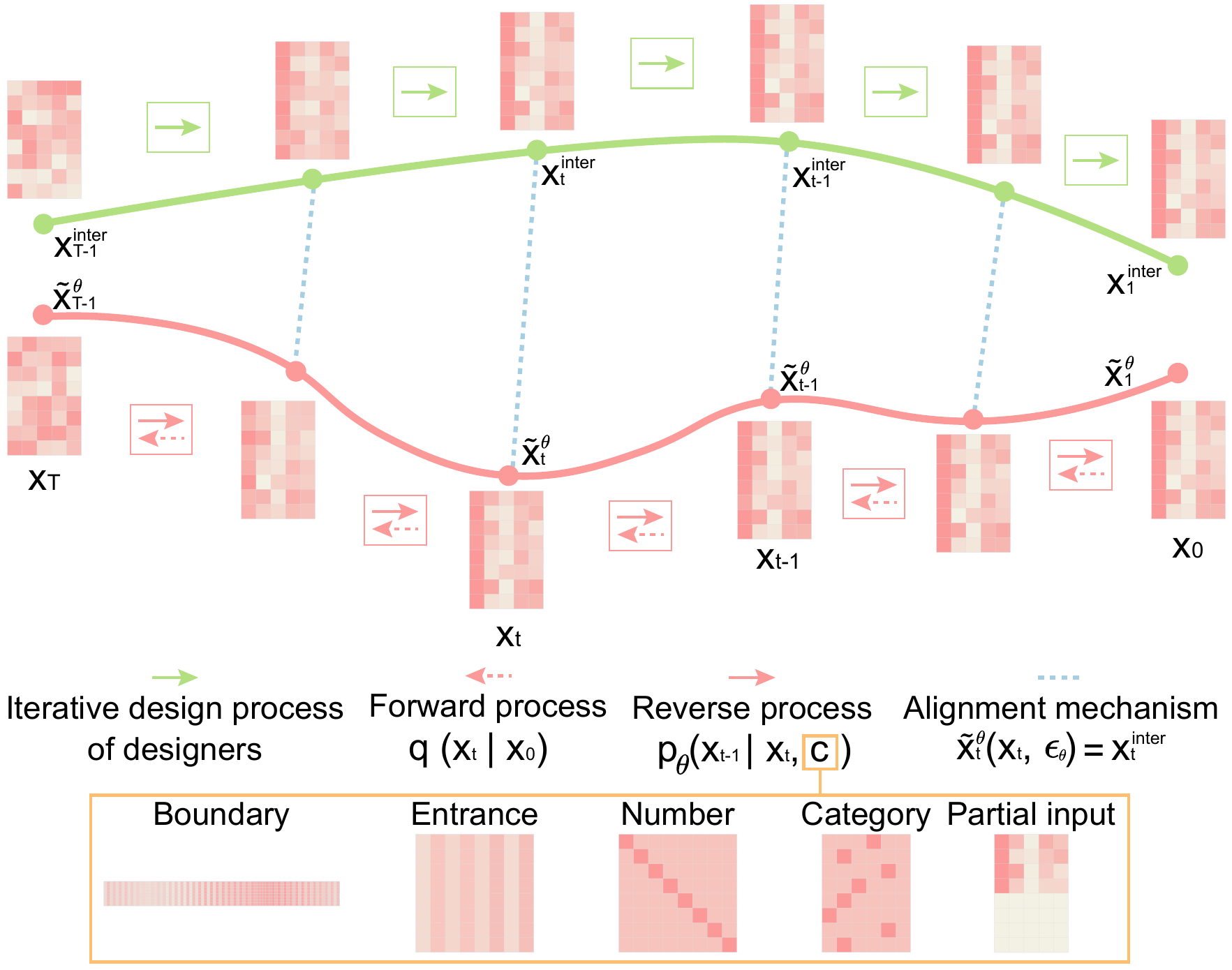}
    \caption{Architecture of NodeDiff. NodeDiff is a Transformer-based conditional diffusion model that integrates an alignment mechanism during training, aligning the optimization trajectory of model with the iterative design processes of designers. This enables our model to handle complex vector data, better fit the distribution of the predicted targets, accomplish the challenging task of layout planning, and achieve diverse user-controllable generation.}
    \label{fig:nodediff}
\end{figure}

\emph{NodeDiff.}
We propose NodeDiff, a Transformer~\cite{Vaswani-17}-based conditional diffusion model for predicting room nodes based on constraints such as the boundary with entrance, room number, room categories, and/or partial input, as shown in Figure~\ref{fig:nodediff}.
Given an initial point $x_0$, we obtain an approximate point $x_t$ by sampling from the posterior distribution $q(x_t | x_0)$ (\ref{eq:forward}), and run a forward step with the inverse process $\epsilon_{\theta}(x_t, t, c)$ conditioning on the constraints $c$ to infer the representation $x_{t-1}$ at time $t-1$:

\begin{equation}
    x_{t-1} = \frac{1}{\sqrt{\alpha_t}}(x_t - \frac{1-\alpha_t}{\sqrt{1-\bar \alpha_t}}\epsilon_\theta(x_t, t, c))+\sigma_t z.
    \label{eq:reverse_c}
\end{equation}
where ${\epsilon}_{\theta}$ is optimized by:

\begin{equation}
    \mathcal{L}_{\text{t-1}} = ||{\epsilon}_{\theta} (x_t, t, c) - {\epsilon} ||^2.
    \label{eq:diff_loss_c}
\end{equation}

\emph{Alignment.}
By rewriting (\ref{eq:forward}), we can obtain $\tilde x^{\theta}_{t}$, an approximation of $x_0$ at each time $t$, using the predicted noise ${\epsilon}_{\theta}$ as:

\begin{equation}
	\label{eq:predicted}
	\tilde x^{\theta}_{t} = (x_t - \sqrt{1-\bar{\alpha}_t} {\epsilon}_{\theta}(x_t, t, c)) / \sqrt{\bar{ \alpha_t}}.
\end{equation}
Some recent work~\cite{Austin-21} additionally introduces a reconstruction loss to encourage plausible predictions of $x_0$ at each time step $t$ by computing the Mean Squared Error (MSE) between the $\tilde x^{\theta}_{t}$ and $x_0$.
Inspired by this, we designed an alignment mechanism for NodeDiff.
Layout planning is often an iterative design process where designers gradually add elements to refine the design.
We hypothesize that aligning the sampling trajectory of the diffusion model with the iterative design process of a designer during training can encourage the model to more closely mimic the experience of designers.
This alignment aims to enhance the ability of model to generate coherent and practical design outcomes.

Design is often carried out element by element, so the iterative design process can be defined as a process of gradually adding elements as the time step $t$ decreases ($T \sim 0$).
An intermediate result $x^{inter}_{t}$ in the iterative design process can be represented by elements that have been confirmed by the designer and elements that remain unconfirmed in the mind of the designer.
For a design target $x_0$ with $n$ elements to be predicted, we can easily calculate the number of confirmed $n_{e_c} = min(n (1 - t / T) + 1, n)$ and unconfirmed elements $n_{e_{unc}} = n - n_{e_c}$ in the intermediate result $x^{inter}_{t}$ at time step $t$.
Clearly, the confirmed elements $e_c$ should match the prediction target values, while the unconfirmed elements $e_{unc}$ should also possess a certain level of plausibility, as the designer may already have a rough idea of them, even if they have not been finalized.
Thus, we first initialize $x^{inter}_{t}$ with $x_0$.
Then we randomly select $n_{e_{unc}}$ indices and set the elements with these indices to a weighted combination of $x_0$ and random noise $x_{noise} \sim \mathcal{N}(0,\mathbf{I})$, ${x^{inter}_{t}}_{e_{unc}} = {x_0}_{e_{unc}} k + {x_{noise}}_{e_{unc}} (1-k)$, where $k = (1 + n_{e_c} / n) / 2$.
As the number of confirmed elements $n_{e_c}$ increases, $k$ will approach $1$.
Because, as the design becomes progressively refined, the unconfirmed elements $e_{unc}$ will increasingly resemble the final design $x_0$.

Now, at each time step $t$, we have a rough estimate $\tilde x^{\theta}_{t}$ of $x_0$ and an intermediate result $x^{inter}_{t}$ corresponding to the iterative design process of the designer.
Thus, the total loss is defined as $\mathcal{L} = \mathcal{L}_{\text{t-1}} + \lambda \mathcal{L}_{\text{align}}$:

\begin{equation}
    \mathcal{L} = || {\epsilon}_{\theta} (x_t, t, c) - {\epsilon} ||^2 + \lambda || \tilde x^{\theta}_{t} (x_t, {\epsilon}_{\theta}) - {x^{inter}_{t}} ||^2.
    \label{eq:total_loss}
\end{equation}
where $\mathcal{L}_{\text{t-1}}$ is the conditional DDPM~\cite{Ho-20} loss for time step $t$, and $\mathcal{L}_{\text{align}}$ is the alignment loss for time step $t$.
This alignment mechanism effectively guides the sampling trajectory at each step to match the iterative design process, distilling the optimizer throughout the reverse process in training.

\emph{User-controllable generation.}
Our NodeDiff is not only fully automated but also easily supports user-controllable generation.
By simply modifying the conditions input to the model in (\ref{eq:diff_loss_c}), various forms of user control can be achieved.
NodeDiff supports a range of user inputs, such as specifying whether or not to include the boundary and entrance, providing the desired number of rooms and their categories, and incorporating partial input.
This flexibility allows users to guide the generation process according to their specific needs and preferences.

\emph{Implementation details.}
We implemented our NodeDiff using PyTorch, based on the public implementation of Guided-Diffusion~\cite{Dhariwal-21}.
The optimizer used is Adam~\cite{Kingma-14} with decoupled weight decay~\cite{Loshchilov-18} for $300k$ steps with a batch size of $1024$ on an NVIDIA RTX $2080$ Ti GPU.
The initial learning rate was set to $1e$-$3$ and was reduced by a factor of $10$ after every $100k$ steps.
We set the number of diffusion steps to $1000$ and uniformly sampled $t$ during training.
Please refer to the supplementary materials for more details.

\subsection{Adjacency Generation} \label{subsec:AdjacencyDiff}

\emph{Problem.}
In this task, our goal is to predict room adjacencies from the input boundary with the entrance, as well as the room nodes generated by NodeDiff.
The room adjacencies reflect the relationships between rooms, forming an essential part of the bubble diagram and serving as a crucial condition for the subsequent floor plan design.

\emph{Representation.}
Previous methods represent room adjacencies as specific line segments in raster space to create a connect mask~\cite{Sun-23}, combined with a rasterized node mask, to indicate room adjacencies (Figure~\ref{fig:overview}(c)).
In contrast, we adopt a more intuitive and efficient approach by using an adjacency matrix to represent the adjacencies between rooms.
In this matrix, a value of $1$ indicates that two nodes are connected, while a value of $0$ indicates that they are not.
We then normalize the matrix to a range of $[-1, 1]$.
This new representation allows the prediction of room adjacencies to be carried out entirely in vector space, eliminating the need for any rasterization operations.

\emph{AdjacencyDiff.}
We propose AdjacencyDiff to predict room adjacencies under the conditions of boundary, entrance, room nodes, and/or partial input.
Its network architecture, training strategy, and implementation details are the same as those of NodeDiff.

\emph{Alignment.}
The design of room adjacencies is also an iterative process. 
Therefore, AdjacencyDiff incorporates an alignment strategy in training, with slight differences from NodeDiff in obtaining intermediate results $x^{inter}_{t}$.
We first extract a list containing $n$ room adjacencies to be predicted, where each room adjacency is a tuple of room IDs.
We then calculate the number of room adjacencies $n_{e_c} = min(n (1 - t / T) + 1, n)$ that should be confirmed at time step $t$.
Next, we randomly select $n_{e_c}$ adjacencies as confirmed adjacencies, while the remaining $n_{e_{unc}} = n - n_{e_c}$ adjacencies are considered unconfirmed.
Similar to NodeDiff, we first initialize $x^{inter}_{t}$ as $x_0$.
Then, for the indices represented by the unconfirmed room adjacencies $e_{unc}$, we assign a weighted sum of $x_0$ and random noise $x_{noise} \sim \mathcal{N}(0,\mathbf{I})$, ${x^{inter}_{t}}_{e_{unc}} = {x_0}_{e_{unc}} k + {x_{noise}}_{e_{unc}} (1-k)$, where $k = (1 + n_{e_c} / n) / 2$.

\emph{User-controllable generation.}
Similar to NodeDiff, our AdjacencyDiff model also supports generating room adjacencies under various conditions, such as specifying the boundary or not, and providing partial input.
This can be easily achieved by simply adding or removing conditions in (\ref{eq:diff_loss_c}).

\subsection{Partitioning Generation} \label{subsec:PartitioningDiff}

\emph{Problem.}
In this task, our goal is to predict room partitioning from the input boundary with entrance, as well as the room nodes generated by NodeDiff and the room adjacencies predicted by AdjacencyDiff.
This will generate a floor plan that adapts to the final stage of the design process.

\emph{Representation.}
Unlike some previous methods that represent room partitioning as a set of multi-channel binary raster masks~\cite{Sun-22, Wu-19} or three-channel color images~\cite{Wang-21, Chaillou-20}, we follow the approach in~\cite{Hu-20, He-22} by using the coordinates of the top-left and bottom-right corners of room bounding boxes to represent room layouts.
This allows our model to perform the prediction process entirely in vector space.

\emph{PartitioningDiff \& Alignment.}
We propose PartitioningDiff to predict room partitioning under the conditions of boundary, entrance, nodes, adjacencies, and/or partial input.
Its network architecture, training strategy, alignment mechanism, and implementation details are the same as those of NodeDiff.

\emph{User-controllable generation.}
By simply adding or removing conditions in (\ref{eq:diff_loss_c}), our PartitioningDiff also supports generating room partitioning under various conditions, such as specifying the boundary or not, and providing partial input.
\section{Evaluation and Discussion}

We conduct quantitative comparisons, qualitative evaluations, ablation experiments, and perceptual studies to thoroughly evaluate our method, concluding with a detailed discussion at the end.

\begin{table}[t]
    \centering
    \caption{Quantitative comparison of FID scores for bubble diagram generation between our method and its variants supporting different types of user inputs (DiffPlanner), and the state-of-the-art method (BubbleFormer).}
    \label{tab:quantitative_bubble}
    \begin{tabular}{|l|cccccc|c|}
        \hline
        Bubble Diagram & \multicolumn{5}{c}{Input Condition} & Output & FID \\
        Generation & $B$ & $R_n$ & $R_c$  & $R_s$  & $R_l$  & w/~Proc & Score \\
        \hline
        BubbleFormer &\cmark  & \xmark & \xmark & \xmark & \xmark & \cmark  & 14.90  \\
        $\text{DiffPlanner}^{b}_{\text{I}}$ & \cmark & \xmark & \xmark & \xmark & \xmark & \xmark  & 1.59 \\
        $\text{DiffPlanner}^{b}_{\text{II}}$ & \cmark & \cmark & \xmark & \xmark & \xmark & \xmark  & 1.39\\
        $\text{DiffPlanner}^{b}_{\text{III}}$ & \cmark & \cmark & \cmark & \xmark & \xmark & \xmark  & 1.19\\
        $\text{DiffPlanner}^{b}_{\text{IV}}$ & \cmark & \cmark & \cmark & \cmark & \cmark & \xmark  & \textbf{0.05} \\
        \hline
        $\text{DiffPlanner}_{\text{I}}$ & \xmark & \xmark & \xmark & \xmark & \xmark & \xmark  & 0.66\\
        $\text{DiffPlanner}_{\text{II}}$ & \xmark & \cmark & \xmark & \xmark & \xmark & \xmark  & 0.64\\
        $\text{DiffPlanner}_{\text{III}}$ & \xmark & \cmark & \cmark & \xmark & \xmark & \xmark  & 0.59\\
        $\text{DiffPlanner}_{\text{IV}}$ & \xmark & \cmark & \cmark & \cmark & \cmark & \xmark  & \textbf{0.05} \\
        \hline
    \end{tabular}
\end{table}

\subsection{Quantitative Comparisons} \label{subsec:quantitative}

\subsubsection{Method variants}

To comprehensively evaluate the performance of our DiffPlanner in generating bubble diagrams and floor plans under various input conditions, we define multiple variants based on different input settings:

\begin{itemize}
    \item 
    $\text{DiffPlanner}^{b}_{\text{I}}$: Generates bubble diagrams and floor plans using only the input boundary with entrance ($B$), without requiring any additional conditions.
    \item 
    $\text{DiffPlanner}^{b}_{\text{II}}$: Extends $\text{DiffPlanner}^{b}_{\text{I}}$ by incorporating the number of rooms ($R_n$) as an additional input condition.
    \item 
    $\text{DiffPlanner}^{b}_{\text{III}}$: Further extends $\text{DiffPlanner}^{b}_{\text{II}}$ by including room categories ($R_c$) as an additional input condition.
    \item 
    $\text{DiffPlanner}^{b}_{\text{IV}}$: Builds upon $\text{DiffPlanner}^{b}_{\text{III}}$ by adding room sizes ($R_s$) and locations ($R_l$) as input conditions.
    \item 
    $\text{DiffPlanner}^{b}_{\text{V}}$: Expands $\text{DiffPlanner}^{b}_{\text{IV}}$ by additionally incorporating room adjacencies ($R_a$) as an input condition, enabling the model to predict floor plans from given bubble diagrams.
\end{itemize}

For the boundary-unconstrained generation of bubble diagrams and floor plans, we define the variants ($\text{DiffPlanner}_{\text{I}}$, \dots, $\text{DiffPlanner}_{\text{V}}$) with the same setup as above, except that the input conditions do not include the boundary with entrance ($B$).

\subsubsection{FID comparison}

A quantitative evaluation of generated bubble diagrams and floor plans is challenging.
Nevertheless, we compare our approach to state-of-the-art techniques using the Fréchet Inception Distance (FID) score~\cite{Heusel-17}, as most state-of-the-art methods~\cite{Sun-22, He-22, Sun-23, Shabani-23} utilized for evaluation.
The FID score is a global metric to calculate the distribution similarity between real data and generated data.
The lower the FID score, the more similar the generated data to the ground truths.
It is important to note that the FID score is highly correlated with the number of examples.
We evaluate each method with different constraints on a test dataset containing $12K$ examples and calculate the respective FID scores.

\emph{Bubble diagram generation.}
We use FID scores to compare our method and its variants supporting different types of user inputs ($\text{DiffPlanner}^{b}_{\text{I, \dots, IV}}$), with the most recent state-of-the-art method~\cite{Sun-23} (BubbleFormer).
Table~\ref{tab:quantitative_bubble} shows that our method ($\text{DiffPlanner}^{b}_{\text{I, \dots, IV}}$) significantly outperforms BubbleFormer in generating bubble diagrams under boundary constraints.
Moreover, unlike BubbleFormer, which only supports fully automated bubble diagram generation without any user input, our method allows for user control over a range of conditions, including the number ($R_n$), categories ($R_c$), sizes ($R_s$), locations ($R_l$) of room nodes.
Note that, BubbleFormer needs to employ the Hungarian matching algorithm~\cite{Kuhn-55} to obtain unique matches between the generated nodes and edges.
While our method does not require any post-processing (Proc) operations.

Our method also supports user-controllable bubble diagram generation without boundary constraints ($\text{DiffPlanner}_{\text{I, \dots, IV}}$), and as more user inputs are incorporated, the performance of the model improves, achieving lower FID scores.
The low FID scores in Table~\ref{tab:quantitative_bubble} clearly demonstrate the ability of our method to accommodate various user conditions and generate high-quality bubble diagrams that closely resemble the ground truths, even without boundary constraints.

\begin{table}[t]
    \centering
    \caption{Quantitative comparison of FID scores for floor plan generation between our method (DiffPlanner), state-of-the-art methods (ArchiGAN, RPLAN, ActFloor-GAN, WallPlan, iPLAN, Graph2Plan, and HouseDiffusion), and their various variants under different conditions.}
    \label{tab:quantitative_fp}
    \resizebox{\columnwidth}{!}{\begin{tabular}{|l|ccccccc|c|}
        \hline
        Floor Plan & \multicolumn{6}{c}{Input Condition} & Output & FID \\
        Generation & $B$ & $R_n$ & $R_c$  & $R_s$  & $R_l$  & $R_a$  & w/ Proc & Score \\
        \hline
        ArchiGAN &\cmark  & \xmark & \xmark & \xmark & \xmark & \xmark & \xmark & 117.59  \\
        RPLAN & \cmark & \xmark & \xmark & \xmark & \xmark & \xmark & \cmark & 4.64 \\
        ActFloor-GAN & \cmark & \xmark & \xmark & \xmark & \xmark & \xmark & \cmark & 4.54 \\
        WallPlan & \cmark & \xmark & \xmark & \xmark & \xmark & \xmark & \cmark & 2.55 \\
        $\text{iPLAN}_{\text{I}}$ & \cmark & \xmark & \xmark & \xmark & \xmark & \xmark & \cmark & 5.38 \\
        $\text{iPLAN}_{\text{II}}$ & \cmark & \cmark & \cmark & \xmark & \xmark & \xmark & \cmark & 5.21 \\
        $\text{iPLAN}_{\text{III}}$ & \cmark & \cmark & \cmark & \xmark & \cmark & \xmark & \cmark & 2.13 \\
        $\text{Graph2Plan}_{\text{I}}$ & \cmark & \xmark & \xmark & \xmark & \xmark & \xmark & \cmark & 2.03 \\
        $\text{Graph2Plan}_{\text{II}}$ & \cmark & \cmark & \cmark & \cmark & \cmark & \cmark & \cmark & 0.50 \\
        $\text{Graph2Plan}_{\text{III}}$ & \cmark & \cmark & \cmark & \cmark & \cmark & \cmark & \xmark & 4.84 \\
        $\text{DiffPlanner}^{b}_{\text{I}}$ & \cmark & \xmark & \xmark & \xmark & \xmark & \xmark & \cmark & 1.08 \\
        $\text{DiffPlanner}^{b}_{\text{II}}$ & \cmark & \cmark & \xmark & \xmark & \xmark & \xmark & \cmark & 0.93\\
        $\text{DiffPlanner}^{b}_{\text{III}}$ & \cmark & \cmark & \cmark & \xmark & \xmark & \xmark & \cmark & 0.72\\
        $\text{DiffPlanner}^{b}_{\text{IV}}$ & \cmark & \cmark & \cmark & \cmark & \cmark & \xmark & \cmark & 0.26 \\
        $\text{DiffPlanner}^{b}_{\text{V}}$ & \cmark & \cmark & \cmark & \cmark & \cmark & \cmark & \cmark & \textbf{0.25} \\
        $\text{DiffPlanner}^{b}_{\text{VI}}$ & \cmark & \cmark & \cmark & \cmark & \cmark & \cmark & \xmark & 0.86 \\
        \hline
        HouseDiffusion &\xmark  & \cmark & \cmark & \xmark & \xmark & \cmark & \xmark & 29.56  \\
        $\text{DiffPlanner}_{\text{I}}$ & \xmark & \xmark & \xmark & \xmark & \xmark & \xmark & \cmark &3.38\\
        $\text{DiffPlanner}_{\text{II}}$ & \xmark & \cmark & \xmark & \xmark & \xmark & \xmark & \cmark &3.25\\
        $\text{DiffPlanner}_{\text{III}}$ & \xmark & \cmark & \cmark & \xmark & \xmark & \xmark & \cmark &3.49\\
        $\text{DiffPlanner}_{\text{IV}}$ & \xmark & \cmark & \cmark & \cmark & \cmark & \xmark & \cmark & 2.89 \\
        $\text{DiffPlanner}_{\text{V}}$ & \xmark & \cmark & \cmark & \cmark & \cmark & \cmark & \cmark & \textbf{2.86}\\
        \hline
    \end{tabular}}
\end{table}

\emph{Floor plan generation.}
We conduct a quantitative comparison of FID scores between our method (DiffPlanner), the state-of-the-art methods (ArchiGAN~\cite{Chaillou-20}, RPLAN~\cite{Wu-19}, ActFloor-GAN~\cite{Wang-21}, WallPlan~\cite{Sun-22}, iPLAN~\cite{He-22}, Graph2Plan~\cite{Hu-20}, \& HouseDiffusion~\cite{Shabani-23}), and their various variants under different conditions including boundary with entrance ($B$), number ($R_n$), categories ($R_c$), sizes ($R_s$), locations ($R_l$), and adjacencies ($R_a$) of rooms, as shown in Table~\ref{tab:quantitative_fp}.

For the boundary-constrained floor plan generation task, our method supports the most user conditions and achieves the lowest FID scores, as shown in Table~\ref{tab:quantitative_fp}.
The ArchiGAN, based on Pix2Pix~\cite{Isola-17}, directly predicts the raster floor plan from the input raster boundary, resulting in excessively noisy outputs that cannot be vectorized, thus obtaining high FID score.
The RPLAN, ActFloor-GAN, and WallPlan use location masks, activity maps, and wall maps as intermediate prediction targets before further predicting the floor plans, resulting in comparatively lower and better FID scores.
Similar to our approach, iPLAN decomposes the floor plan generation process, using a set of node masks that includes room number, categories, and locations as the intermediate representation.
This allows their models to support more user conditions.
Both Graph2Plan and our method employ a boundary and bubble diagram-driven layout generation strategy.
We also employ a post-processing (Proc) step as in Graph2Plan to eliminate gaps that may exist between the predicted rooms.
Eventually, our method consistently achieves lower FID scores, both with and without post-processing, surpassing Graph2Plan.

For the task of floor plan generation without boundary constraints, previous methods~\cite{Nauata-20, Nauata-21, Shabani-23} often support only a single user condition, typically a bubble diagram-constrained floor plan generation.
Compared to the previous state-of-the-art method, HouseDiffusion~\cite{Shabani-23}, our method supports a wider range of user inputs and achieves significantly lower FID scores.
This indicates that our method is more flexible and capable of generating higher-quality floor plans that closely resemble the ground truths.

\begin{table}[t]
    \centering
    \caption{Statistics comparison. Each statistic is the ratio calculated based on the ground truth bubble diagrams and floor plans. A ratio close to $1$ suggests that the geometry and topology of the model-generated results are more similar to the ground truths.}
    \label{tab:quantitative_fp_statistics}
    \begin{tabular}{|l|ccccc|}
        \hline
        Bubble Diagram & \multicolumn{5}{c|}{Statistics}\\
        Generation & $R^n_{avg}$  & $C^l_{avg}$  & $C^r_{avg}$ & $L^n_{avg}$ & $L^a_{avg}$  \\
        \hline
        BubbleFormer & 0.932 & 0.376 & 0.369 & 0.969 & 1.074  \\
        $\text{DiffPlanner}^{b}_{\text{I}}$ & \textbf{0.996} & \textbf{1.000} & \textbf{1.004} & \textbf{0.998} & \textbf{1.004} \\
        \hline
        Floor Plan & \multicolumn{5}{c|}{Statistics}\\
        Generation & $R^n_{avg}$  & $C^l_{avg}$  & $C^r_{avg}$ & $L^n_{avg}$ & $L^a_{avg}$  \\
        \hline
        RPLAN & 0.869 & 0.851 & 0.997 & 0.958 & 1.045  \\
        ActFloor-GAN & 0.904 & 0.864 & 0.965 & 0.935 & 0.902 \\
        WallPlan    & \textbf{0.998} & 0.968 & 0.973 & 0.970 & 0.911 \\
        $\text{iPLAN}_{\text{I}}$ & 0.938 & 0.980 & 1.049 & 0.969 & 1.131 \\
        $\text{Graph2Plan}_{\text{I}}$ & 0.980 & 0.971 & 0.988 & 0.969 & \textbf{1.029} \\
        $\text{DiffPlanner}^{b}_{\text{I}}$ & 0.996 & \textbf{0.997} & \textbf{1.001} & \textbf{0.998} & \textbf{0.971} \\
        \hline
    \end{tabular}
\end{table}

\subsubsection{Statistics comparsion}

FID is a metric that assesses the similarity in overall distribution between generated results and real data.
However, it fails to specifically account for the intricate geometric and topological details in the quality of generated bubble diagrams and floor plans.
Similar to~\cite{Sun-22, Sun-23}, we have gathered various statistical metrics to evaluate the geometry and topology of these generated bubble diagrams and floor plans.

These statistics include the total number of rooms ($R^n$), the number of rooms directly connected to the living room ($C^l$), the ratio of rooms connected to the living room to all non-living rooms ($C^r$), the total number of living rooms ($L^n$), and the percentage of area occupied by the living room ($L^a$).
We compute the average for each statistical metric across the test dataset containing $12K$ examples and calculate its ratio with the corresponding data from the ground truths.
A ratio close to $1$ suggests that the geometry and topology of the model-generated results are more similar to the ground truths.

We conduct a statistics comparison between our method ($\text{DiffPlanner}^{b}_{\text{I}}$) and the state-of-the-art methods (BubbleFormer~\cite{Sun-23}, RPLAN~\cite{Wu-19}, ActFloor-GAN~\cite{Wang-21}, WallPlan~\cite{Sun-22}, $\text{iPLAN}_{\text{I}}$~\cite{He-22}, \& $\text{Graph2Plan}_{\text{I}}$~\cite{Hu-20}), focusing on the tasks of generating bubble diagrams and floor plans with boundaries as the only condition.
This comparison is intended to evaluate the geometric and topological characteristics of the model-generated results.

The results of the comparison are presented in Table~\ref{tab:quantitative_fp_statistics}.
Compared to the state-of-the-art method in bubble diagram generation, BubbleFormer, our method surpasses it in all five metrics, particularly excelling in the two statistics related to room adjacencies ($C^l$ \& $C^r$).
In the task of floor plan generation, our method achieves the best performance in four out of five metrics compared to the existing methods ($C^l$, $C^r$, $L^n$ \& $L^a$).
The results thus indicate that our method can generate outputs that most closely resemble the ground truth geometry and topology, surpassing the existing state-of-the-art methods.

\begin{table}[t]
    \centering
    \caption{Quantitative evaluation of how well our DiffPlanner adheres to the input conditions under different control modes for bubble diagram and floor plan generation tasks. A lower $\text{MAE}_{avg}$ indicates stricter adherence to the input conditions.}
    \label{tab:quantitative_cond_compliance}
    \resizebox{\columnwidth}{!}{\begin{tabular}{|l|ccc|cccc|}
        \hline
        Bubble Diagram & \multicolumn{2}{c}{Input Condition} & Output & \multicolumn{4}{c|}{$\text{MAE}_{avg}$} \\
        Generation & $B$ & $R_{n,c,s,l}$ & w/~Proc  & $R_{n}$  & \multicolumn{2}{c}{$R_{c}$}  & $R_{s,l}$ \\
        \hline
        $\text{DiffPlanner}^{b}_{\text{II}}$    & \cmark & $R_{n}$         & \xmark & 0.0 &\multicolumn{2}{c}{-}    & - \\
        $\text{DiffPlanner}^{b}_{\text{III}}$   & \cmark & $R_{n,c}$       & \xmark & 0.0 &\multicolumn{2}{c}{0.0}  & - \\
        $\text{DiffPlanner}^{b}_{\text{IV}}$    & \cmark & $R_{n,c,s,l}$   & \xmark & 0.0 &\multicolumn{2}{c}{0.0}  & 0.0 \\
        \hline
        Floor Plan & \multicolumn{2}{c}{Input Condition} & Output & \multicolumn{4}{c|}{$\text{MAE}_{avg}$} \\
        Generation & $B$ & $R_{n,c,s,l,a}$ & w/~Proc  & $R_{n}$  & $R_{c}$  & $R_{s,l}$ & $R_{a}$ \\
        \hline
        $\text{DiffPlanner}^{b}_{\text{II}}$    & \cmark & $R_{n}$         & \cmark & 0.0 & -    & -      & - \\
        $\text{DiffPlanner}^{b}_{\text{III}}$   & \cmark & $R_{n,c}$       & \cmark & 0.0 & 0.0  & -      & - \\
        $\text{DiffPlanner}^{b}_{\text{IV}}$    & \cmark & $R_{n,c,s,l}$   & \cmark & 0.0 & 0.0  & 0.008  & - \\
        $\text{DiffPlanner}^{b}_{\text{V}}$     & \cmark & $R_{n,c,s,l,a}$ & \cmark & 0.0 & 0.0  & 0.008  & 0.025 \\
        $\text{DiffPlanner}^{b}_{\text{VI}}$    & \cmark & $R_{n,c,s,l,a}$ & \xmark & 0.0 & 0.0  & 0.001  & 0.007 \\
        \hline
    \end{tabular}}
\end{table}

\subsubsection{Assessing controllability compliance}

To effectively demonstrate the controllability of our method, we further examine whether the generated results adhere well to the specified input conditions, as shown in Table~\ref{tab:quantitative_cond_compliance}.
We first extract key attributes from the vector outputs of our DiffPlanner, including the number ($R_n$), categories ($R_c$), sizes, locations ($R_{s,l}$), and adjacencies ($R_a$) of rooms.
Specifically, the $R_n$ is represented as an integer, while the $R_c$, $R_{s,l}$, and $R_a$ are represented as $8 \times 1$, $8 \times 3$, \& $8 \times 8$ tensors, respectively, similar to the tensors of nodes and adjacencies illustrated in Figure~\ref{fig:overview}(b).
Then, for each input condition ($R_n$, $R_c$, $R_{s,l}$, \& $R_a$), we compute the Mean Absolute Error (MAE) between the generated result and the input for each sample in the test dataset, which contains $12K$ examples, and report the average ($\text{MAE}_{avg}$).
A lower $\text{MAE}_{avg}$ indicates that the model adheres more strictly to the input conditions.

Table~\ref{tab:quantitative_cond_compliance} shows that our DiffPlanner achieves consistently low $\text{MAE}_{avg}$ across various control modes for boundary-constrained bubble diagram and floor plan generation tasks.
This demonstrates that our method effectively adheres to the input conditions, validating the effectiveness of its controllability.
Specifically, for input conditions $R_n$ and $R_c$, our DiffPlanner does not modify them during the prediction process; they are directly preserved as part of the final output.
For the input conditions $R_{s,l}$, they remains unchanged when predicting the bubble diagram and are directly carried over to the final output.
However, when predicting the floor plan, the final output consists of room boxes inferred by the model based on the input conditions and further refined through a post-processing (Proc) step.
This introduces slight variations in the recalculated $R_{s,l}$ and $R_a$, as they are derived from the predicted room boxes, compared to the input $R_{s,l}$ and $R_a$.
To further analyze this effect, we conduct an ablation study by removing the post-processing step ($\text{DiffPlanner}^{b}_{\text{V}}$ vs. $\text{DiffPlanner}^{b}_{\text{VI}}$).
The results show that the slight variations in $R_{s,l}$ and $R_a$ are primarily caused by the post-processing module, while our model itself still adheres well to the input conditions.

\begin{table}[t]
    \centering
    \caption{Quantitative evaluation of the output diversity of our method ($\text{DiffPlanner}^{b}_{\text{I}}$) on the areas of six room categories, including the living room ($R_{liv}$), bedroom ($R_{bed}$), kitchen ($R_{kit}$), bathroom ($R_{bat}$), balcony ($R_{bal}$), and storage ($R_{sto}$). The $\text{Diversity}_{avg}$ and $\text{Coverage}^{GT}_{avg}$ range from $0$ to $1$, where lower scores indicate higher output diversity.}
    \label{tab:quantitative_diversity}
    \begin{tabular}{|l|cccccc|}
        \hline
        & \multicolumn{6}{c|}{$\text{Diversity}_{avg}$}\\
        & $R_{liv}$  & $R_{bed}$  & $R_{kit}$ &$R_{bat}$ & $R_{bal}$ & $R_{sto}$  \\
        \hline
        $\text{DiffPlanner}^{b}_{\text{I}}$ & 0.57 & 0.53 & 0.26  & 0.19  & 0.34 & 0.00 \\
        \hline
        & \multicolumn{6}{c|}{$\text{Coverage}^{GT}_{avg}$}\\
        & $R_{liv}$  & $R_{bed}$  & $R_{kit}$ &$R_{bat}$ & $R_{bal}$ & $R_{sto}$  \\
        \hline
        $\text{DiffPlanner}^{b}_{\text{I}}$ & 0.56  & 0.52  & 0.24 & 0.17 & 0.34 & 0.00 \\
        \hline
    \end{tabular}
\end{table}

\subsubsection{Output diversity}

We further quantitatively evaluate the diversity of the results generated by our model to assess whether it can produce diverse floor plans given the same input condition.
First, for each sample ($s$) in the test dataset containing $12K$ examples ($S$), we let $\text{DiffPlanner}^{b}_{\text{I}}$ generate five floor plan variants ($V_{i=1, \dots, K, K = 5}$) using only the boundary as input condition.
Then, to measure the similarity between two variants $V_i$ and $V_j$, we compute the Intersection over Union (IoU) separately for the total area of each ($r$) of the six room categories ($R$), including the living room ($R_{liv}$), bedroom ($R_{bed}$), kitchen ($R_{kit}$), bathroom ($R_{bat}$), balcony ($R_{bal}$), and storage ($R_{sto}$), between the corresponding variants ($V_i$ \& $V_j$).
For each sample ($s$), we compute the mean IoU across all pairwise comparisons among the five variants.
Finally, we take the average of these values over the entire test dataset to obtain the following metric:

\begin{equation}
    \text{Diversity}_{avg} = \left[ \mathbb{E}_{s \in S} \mathbb{E}_{1 \leq i < j \leq K} \text{IoU}_{s,r}(V_i,V_j) \right]_{r \in R}.
    \label{eq:diversity_avg}
\end{equation}

The $\text{Diversity}_{avg}$ ranges from $0$ to $1$.
A lower value indicates a smaller IoU between the generated results, implying greater output diversity of the model.
As shown in Table~\ref{tab:quantitative_diversity}, our method performs well on $\text{Diversity}_{avg}$, achieving relatively low scores across different room categories.
This indicates that our method generates outputs with high diversity.
For the living room and bedroom, the scores are $0.57$ and $0.53$, respectively, which are slightly higher than those for other room categories.
This can be expected, as the living room and bedroom typically occupy the largest portions of indoor space in residential buildings.
Additionally, their approximate locations are often constrained by the input boundary and entrance, leading to a certain degree of similarity across different layout variations.
However, since the scores lie in the mid-range between $0$ and $1$, this already represents a strong diversity performance.
Similarly, compared to the kitchen and bathroom, the balcony has more spatial constraints, as it is often positioned along specific protruding edges of the boundary.
As a result, it exhibits a higher score than the kitchen and bathroom.

\begin{table}[t]
    \centering
    \caption{Quantitative evaluation of the ability of our three core components (NodeDiff, AdjacencyDiff, and PartitioningDiff) to predict the complete target from partial inputs under different conditions. The table shows FID scores for each component with varying proportions of target information in the partial inputs: 25\%, 50\%, and 75\%.}
    \label{tab:partial_input}
    \resizebox{\columnwidth}{!}{\begin{tabular}{|l|ccccccc|c|}
        \hline
        & \multicolumn{7}{c|}{Input Condition}& FID \\
        & $B$ & $R_n$ & $R_c$  & $R_s$  & $R_l$  & $R_a$ & Partial & Score \\
        \hline
        $\text{NodeDiff}_{\text{I}}$ & \cmark & \xmark & \xmark & \xmark & \xmark & \xmark & 25\%  & 1.86 \\
        $\text{NodeDiff}_{\text{I}}$ & \cmark & \xmark & \xmark & \xmark & \xmark & \xmark & 50\%  & 0.76\\
        $\text{NodeDiff}_{\text{I}}$ & \cmark & \xmark & \xmark & \xmark & \xmark & \xmark & 75\%  & 0.16\\
        \hline
        $\text{NodeDiff}_{\text{II}}$ & \cmark & \cmark & \xmark & \xmark & \xmark & \xmark & 25\%  & 1.65\\
        $\text{NodeDiff}_{\text{II}}$ & \cmark & \cmark & \xmark & \xmark & \xmark & \xmark & 50\%  &  0.65\\
        $\text{NodeDiff}_{\text{II}}$ & \cmark & \cmark & \xmark & \xmark & \xmark & \xmark & 75\%  &  0.13\\
        \hline
        $\text{NodeDiff}_{\text{III}}$ & \cmark & \cmark & \cmark & \xmark & \xmark & \xmark & 25\%  &  1.13  \\
        $\text{NodeDiff}_{\text{III}}$ & \cmark & \cmark & \cmark & \xmark & \xmark & \xmark & 50\%  &   0.46\\
        $\text{NodeDiff}_{\text{III}}$ & \cmark & \cmark & \cmark & \xmark & \xmark & \xmark & 75\%  & 0.11\\
        \hline
        $\text{AdjacencyDiff}$ & \cmark & \cmark & \cmark & \cmark & \cmark & \xmark & 25\%  &  0.05 \\
        $\text{AdjacencyDiff}$ & \cmark & \cmark & \cmark & \cmark & \cmark & \xmark & 50\%  &  0.05\\
        $\text{AdjacencyDiff}$ & \cmark & \cmark & \cmark & \cmark & \cmark & \xmark & 75\%  & 0.04\\
        \hline
        $\text{PartitioningDiff}$ & \cmark & \cmark & \cmark & \cmark & \cmark & \cmark & 25\%  &  0.76 \\
        $\text{PartitioningDiff}$ & \cmark & \cmark & \cmark & \cmark & \cmark & \cmark & 50\%  & 0.31 \\
        $\text{PartitioningDiff}$ & \cmark & \cmark & \cmark & \cmark & \cmark & \cmark & 75\%  &  0.07 \\
        \hline
    \end{tabular}}
\end{table}

We further analyze the differences between the generated result by our $\text{DiffPlanner}^{b}_{\text{I}}$ (Our) and its corresponding ground truth (GT) for each sample (s) in the test dataset (S), using the following metric:

\begin{equation}
    \text{Coverage}^{GT}_{avg} = \left[ \mathbb{E}_{s \in S} \text{IoU}_{s,r}(Our, GT) \right]_{r \in R}.
    \label{eq:coverage_gt_avg}
\end{equation}

The $\text{Coverage}^{GT}_{avg}$ ranges from $0$ to $1$, where a lower value indicates lower similarity and greater divergence.
Table~\ref{tab:quantitative_diversity} shows that our method achieves a relatively low score on $\text{Coverage}^{GT}_{avg}$, indicating that for each test sample, our model does not simply generate results that closely approximate the GT.
An ideal AI design tool should capture the diverse design space that maps a single input to multiple possible outputs, rather than merely learning a deterministic mapping from a single input to a single output.
Given a boundary as input, the corresponding floor plan design space should be diverse, rather than being limited to the specific GT solution.

The low FID score in Table~\ref{tab:quantitative_fp} suggests that the results generated by our model are globally aligned with the data distribution of GT.
Meanwhile, the relatively low $\text{Coverage}^{GT}_{avg}$ score, which measures the similarity between the generated result and its corresponding GT for each individual sample, further demonstrates that our method does not simply memorize the input-to-single-GT mapping as a fixed solution during training.
Instead, it effectively captures the diverse design space from input to output.
The low $\text{Diversity}_{avg}$ score in Table~\ref{tab:quantitative_diversity} further validates this observation.

\subsubsection{Partial input generation}

We further evaluate the ability of our three core components (NodeDiff, AdjacencyDiff, \& PartitioningDiff) to predict the complete target from partial inputs based on different input conditions including the boundary with entrance ($B$), number ($R_n$), categories ($R_c$), sizes ($R_s$), locations ($R_l$), and adjacencies ($R_a$) of rooms.
We set the proportion of target information in the partial inputs to three values: $25\%$, $50\%$, and $75\%$.

As shown in Table~\ref{tab:partial_input}, even with partial inputs containing only $25\%$ of the target information, all three core components achieve relatively low FID scores.
Furthermore, as the proportion of target information increases, the FID scores further decrease.
These experimental results demonstrate the ability of our method to predict complete layouts from partial inputs, indicating strong support for user interaction.

\begin{figure}[t]
    \centering
    \includegraphics[width=\linewidth]{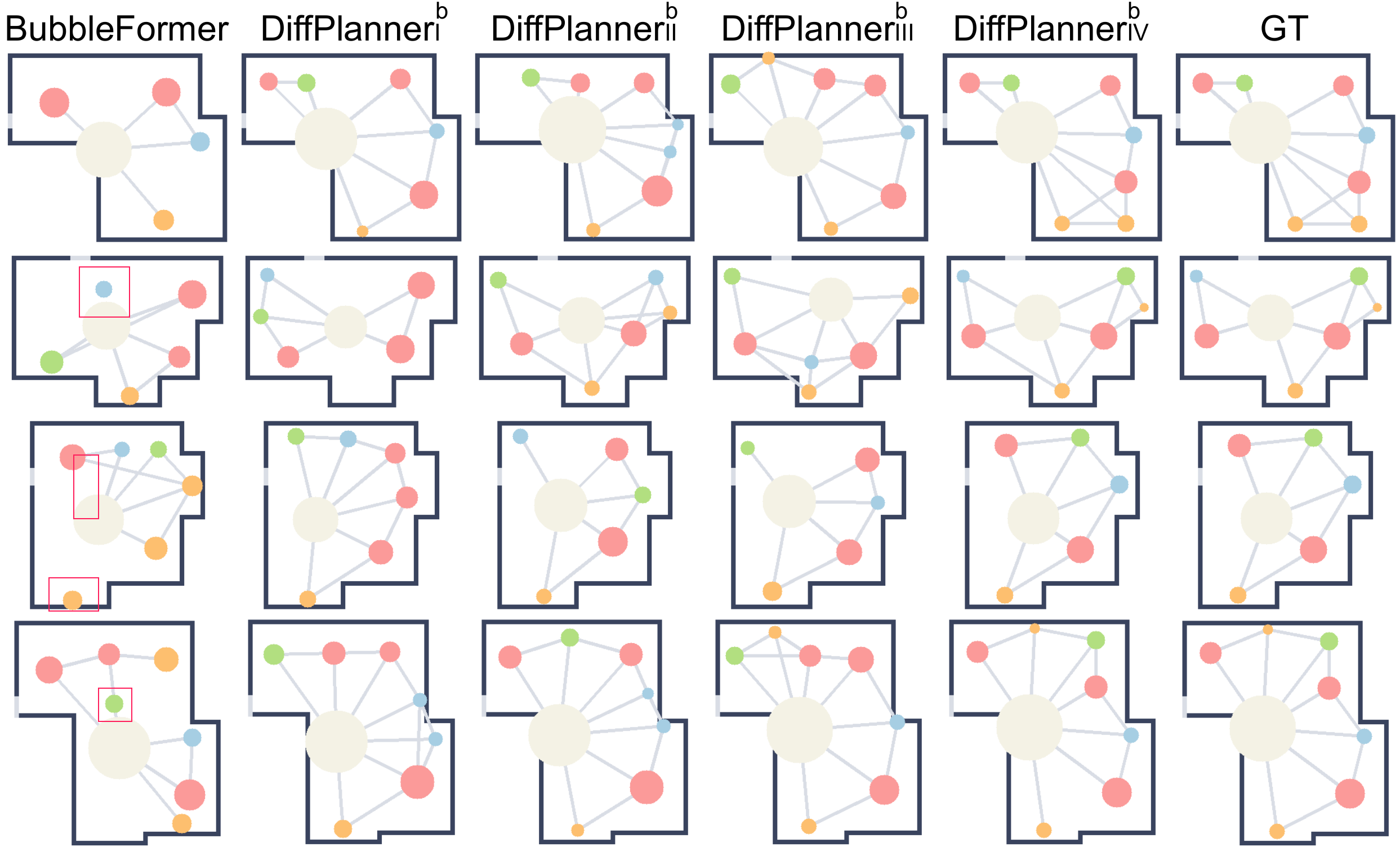}
    \caption{Qualitative comparison between our method based on various conditions ($\text{DiffPlanner}^{b}_{\text{I, \dots, IV}}$) and the state-of-the-art method (BubbleFormer) for boundary-constrained bubble diagram generation. While BubbleFormer can generate bubble diagrams fully automatically, it does not support user interaction. In contrast, our method not only supports fully automatic generation ($\text{DiffPlanner}^{b}_{\text{I}}$) but also allows users to apply various controls, such as the number ($\text{DiffPlanner}^{b}_{\text{II}}$), categories ($\text{DiffPlanner}^{b}_{\text{III}}$), sizes \& locations ($\text{DiffPlanner}^{b}_{\text{IV}}$) of nodes.}
    \label{fig:qualitative_bu_wb}
\end{figure}

\subsection{Qualitative Evaluations} \label{subsec:qualitative}

\subsubsection{Bubble diagram generation}

Figure~\ref{fig:qualitative_bu_wb} presents the qualitative comparison results of our method based on various conditions and the state-of-the-art method BubbleFormer for boundary-constrained bubble diagram generation.
Please note that while BubbleFormer can generate bubble diagrams fully automatically, it does not support user interaction.
In contrast, our method not only supports fully automatic generation ($\text{DiffPlanner}^{b}_{\text{I}}$) but also allows users to apply various controls, such as the number ($\text{DiffPlanner}^{b}_{\text{II}}$), categories ($\text{DiffPlanner}^{b}_{\text{III}}$), and sizes \& locations ($\text{DiffPlanner}^{b}_{\text{IV}}$) of nodes.

BubbleFormer is a VAE-based generative model that first rasterizes room nodes and adjacencies into image space, extracts feature from the rasterized data, and then performs subsequent predictions.
This process often leads to errors in bubble diagram generation, especially in predicting room adjacencies.
Common issues include bathrooms and balconies not adjacent to any other rooms, bedrooms not connected to the living room, and unreasonable kitchen placements, as shown in Figure~\ref{fig:qualitative_bu_wb}.

Compared to BubbleFormer, our fully automated method ($\text{DiffPlanner}^{b}_{\text{I}}$) performs excellently, predicting high-quality and reasonable bubble diagrams without any rasterization operations, as all predictions are made in vector space.
Our method ($\text{DiffPlanner}^{b}_{\text{I, \dots, IV}}$) also supports various user input conditions, and as more user inputs are provided, our results increasingly approximate the ground truths (GT).

\begin{figure}[t]
    \centering
    \includegraphics[width=\linewidth]{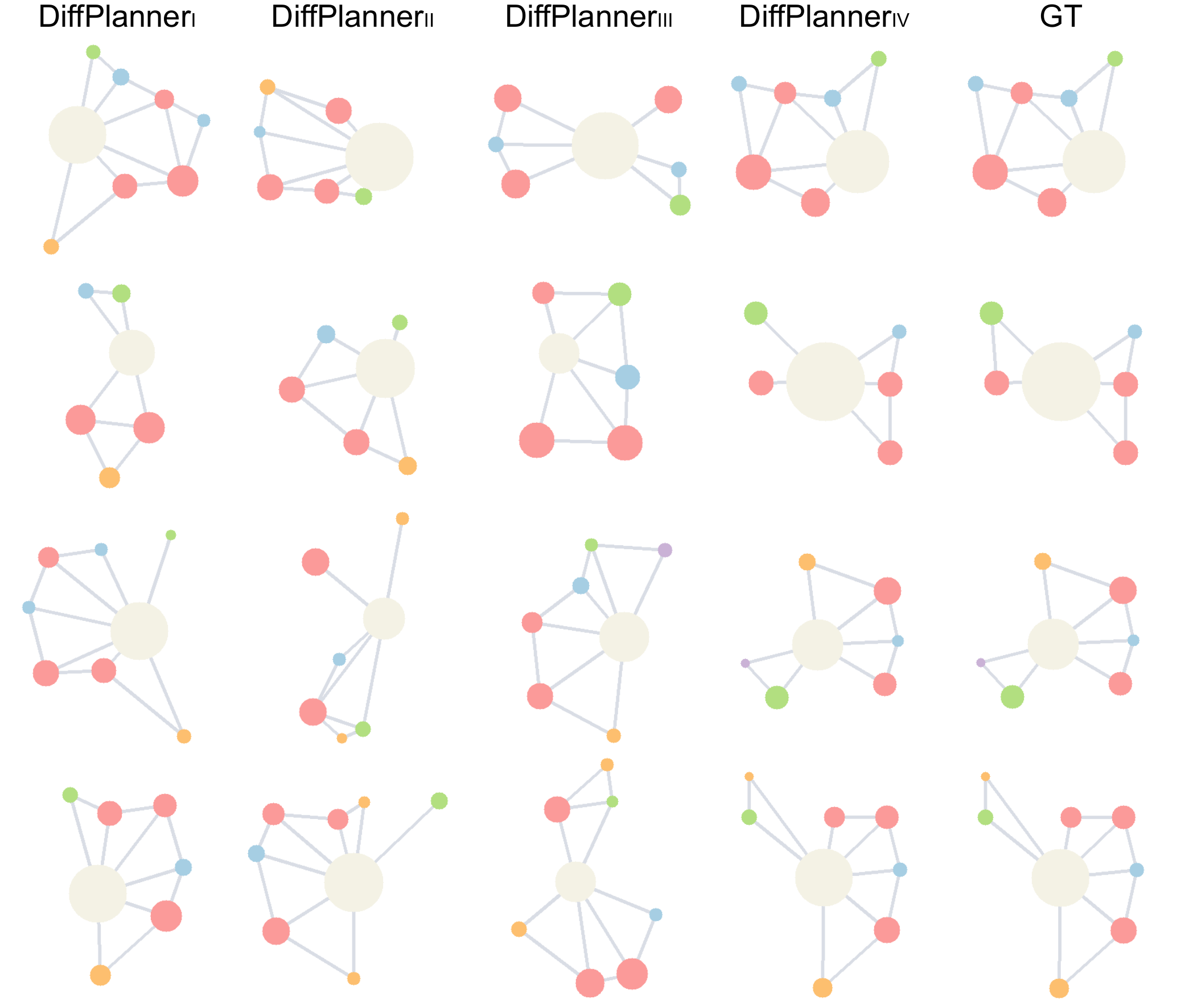}
    \caption{Our method can generate bubble diagrams fully automatically ($\text{DiffPlanner}_{\text{I}}$) or based on various user inputs, such as the number ($\text{DiffPlanner}_{\text{II}}$), categories ($\text{DiffPlanner}_{\text{III}}$), and sizes \& locations ($\text{DiffPlanner}_{\text{IV}}$) of nodes, without boundary constraints.}
    \label{fig:qualitative_bu_wob}
\end{figure}

Figure~\ref{fig:qualitative_bu_wob} demonstrates the ability of our method to generate bubble diagrams fully automatically ($\text{DiffPlanner}_{\text{I}}$) or based on various user inputs, such as the number ($\text{DiffPlanner}_{\text{II}}$), categories ($\text{DiffPlanner}_{\text{III}}$), and sizes \& locations ($\text{DiffPlanner}_{\text{IV}}$) of nodes, without boundary constraints.
This shows that our method is capable of driving downstream boundary-unconstrained layout planning tasks, allowing for the controllable generation of bubble diagrams in the early stages of the design process.

\begin{figure*}[t]
    \centering
    \includegraphics[width=\linewidth]{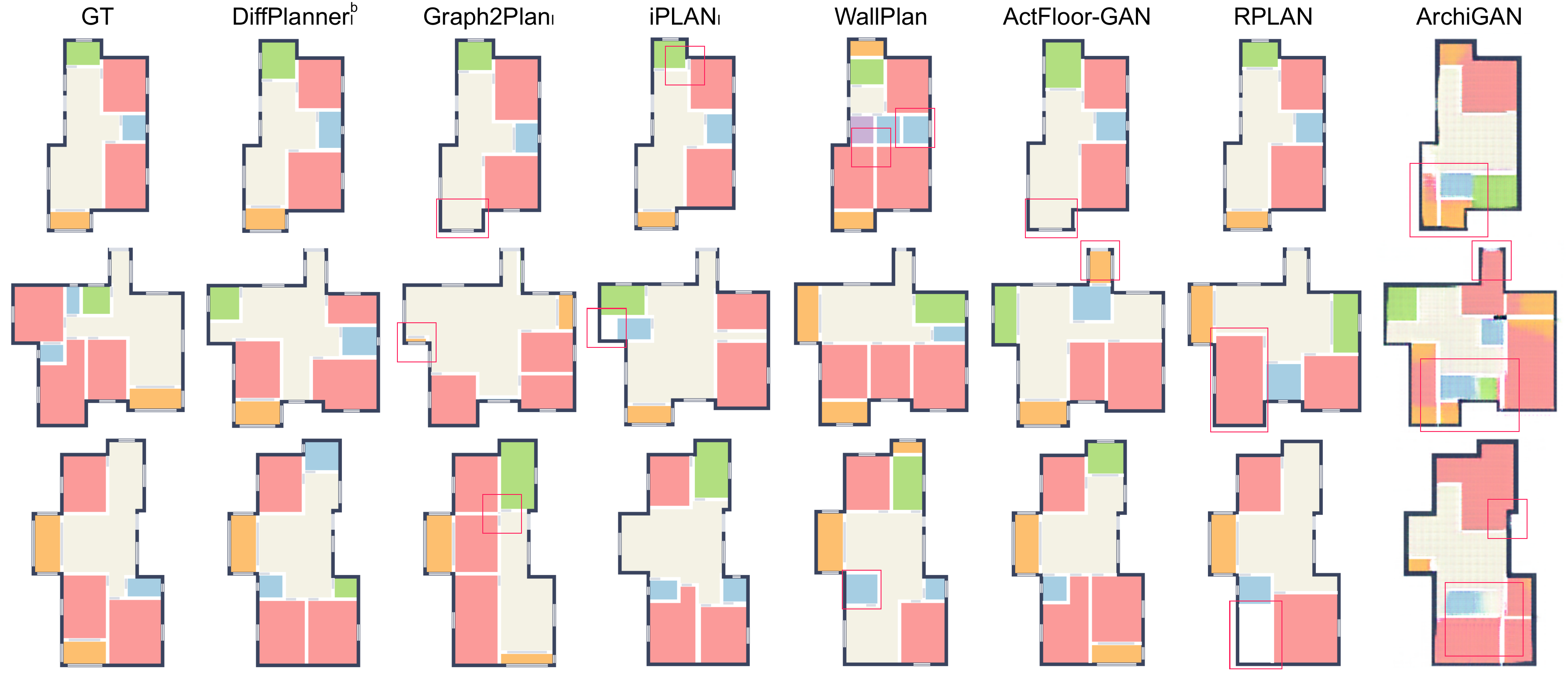}
    \caption{Qualitative comparison of floor plan generation from boundary only between our method ($\text{DiffPlanner}^{b}_{\text{I}}$) and previous state-of-the-art methods ($\text{Graph2Plan}_{\text{I}}$, $\text{iPLAN}_{\text{I}}$, WallPlan, ActFloor-GAN, RPLAN, and ArchiGAN). The unreasonable design is highlighted in the red box. Our method produces floor plans closest to the ground truths (GT), demonstrating the effectiveness of our vector-based approach in generating accurate and high-quality floor plans.}
    \label{fig:qualitative_fp_wb}
\end{figure*}

\begin{figure}[t]
    \centering
    \includegraphics[width=\linewidth]{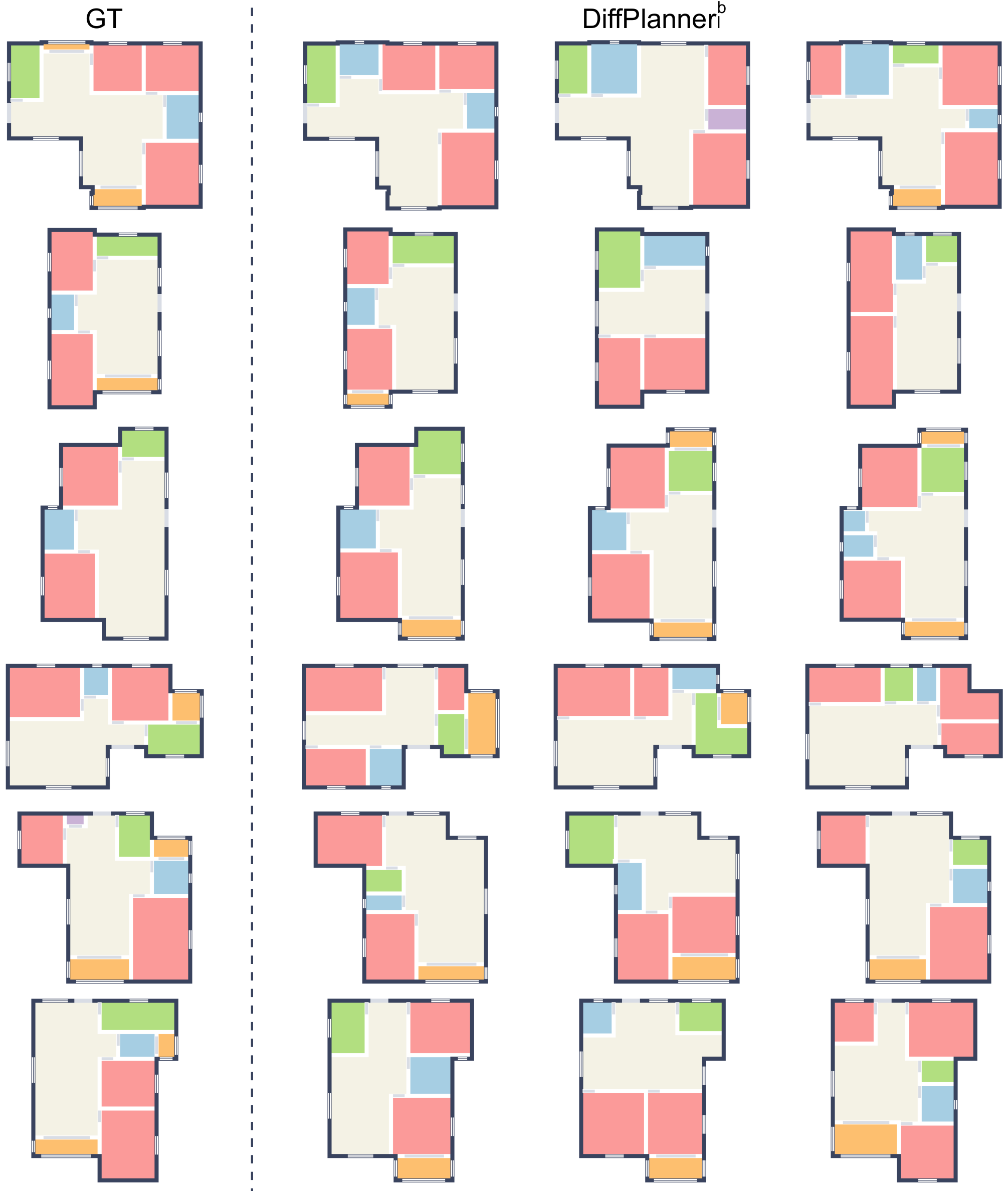}
    \caption{Qualitative evaluation of output diversity. Our method ($\text{DiffPlanner}^{b}_{\text{I}}$) can generate multiple high-quality floor plans from a single input boundary.}
    \label{fig:qualitative_fp_diversity}
\end{figure}

\subsubsection{Floor plan generation}

Figure~\ref{fig:qualitative_fp_wb} shows the qualitative comparison results between our method ($\text{DiffPlanner}^{b}_{\text{I}}$) and previous methods ($\text{Graph2Plan}_{\text{I}}$~\cite{Hu-20}, $\text{iPLAN}_{\text{I}}$~\cite{He-22}, WallPlan~\cite{Sun-22}, ActFloor-GAN~\cite{Wang-21}, RPLAN~\cite{Wu-19}, and ArchiGAN~\cite{Chaillou-20}) for floor plan generation from boundary only.
The door and window placements for all methods used the rule-based algorithm in RPLAN~\cite{Wu-19} and Graph2Plan~\cite{Hu-20}.

ArchiGAN directly predicts rasterized floor plans from rasterized boundaries, resulting in noisy outputs that cannot be vectorized.
Furthermore, methods like RPLAN, ActFloorGAN, and WallPlan start with rasterized boundaries and first predict intermediate information, such as location masks, activity maps, or wall maps, which contain the topological and geometric information of the layouts.
They then perform subsequent predictions, enabling them to generate floor plans that can be vectorized.
Graph2Plan and iPLAN take rasterized boundaries as input but represent their prediction targets as vector bounding box coordinates, although they still require the model to predict a rasterized layout image as an auxiliary.
Unlike all previous methods, our approach operates entirely in vector space, taking vectorized boundaries as input and progressively generating the final high-quality vector floor plans.
As shown in Figure~\ref{fig:qualitative_fp_wb}, our method produces floor plans that are closest to the ground truths (GT), while other methods exhibit varying degrees of unreasonable results.
This demonstrates the effectiveness of our vector-based approach in generating accurate and high-quality floor plans.

Figure~\ref{fig:qualitative_fp_diversity} qualitatively demonstrates that our method ($\text{DiffPlanner}^{b}_{\text{I}}$) can generate multiple different high-quality floor plans from a single input boundary.
This indicates that our model successfully captures the diverse design space in the floor plan generation task, mapping a single input to multiple possible outputs.
Within the diverse outputs of our model, some results may closely resemble the corresponding ground truths (GT), as seen in the samples shown in the second and third rows of Figure~\ref{fig:qualitative_fp_diversity}.
This is expected, as the GT itself is one feasible solution within this diverse design space.
For a given input boundary, we do not expect the outputs of the model to always match the corresponding specific GT.
Instead, we aim for the model to generate multiple reasonable and high-quality floor plans, reflecting the inherent diversity of architectural design solutions.

\begin{figure}[t]
    \centering
    \includegraphics[width=\linewidth]{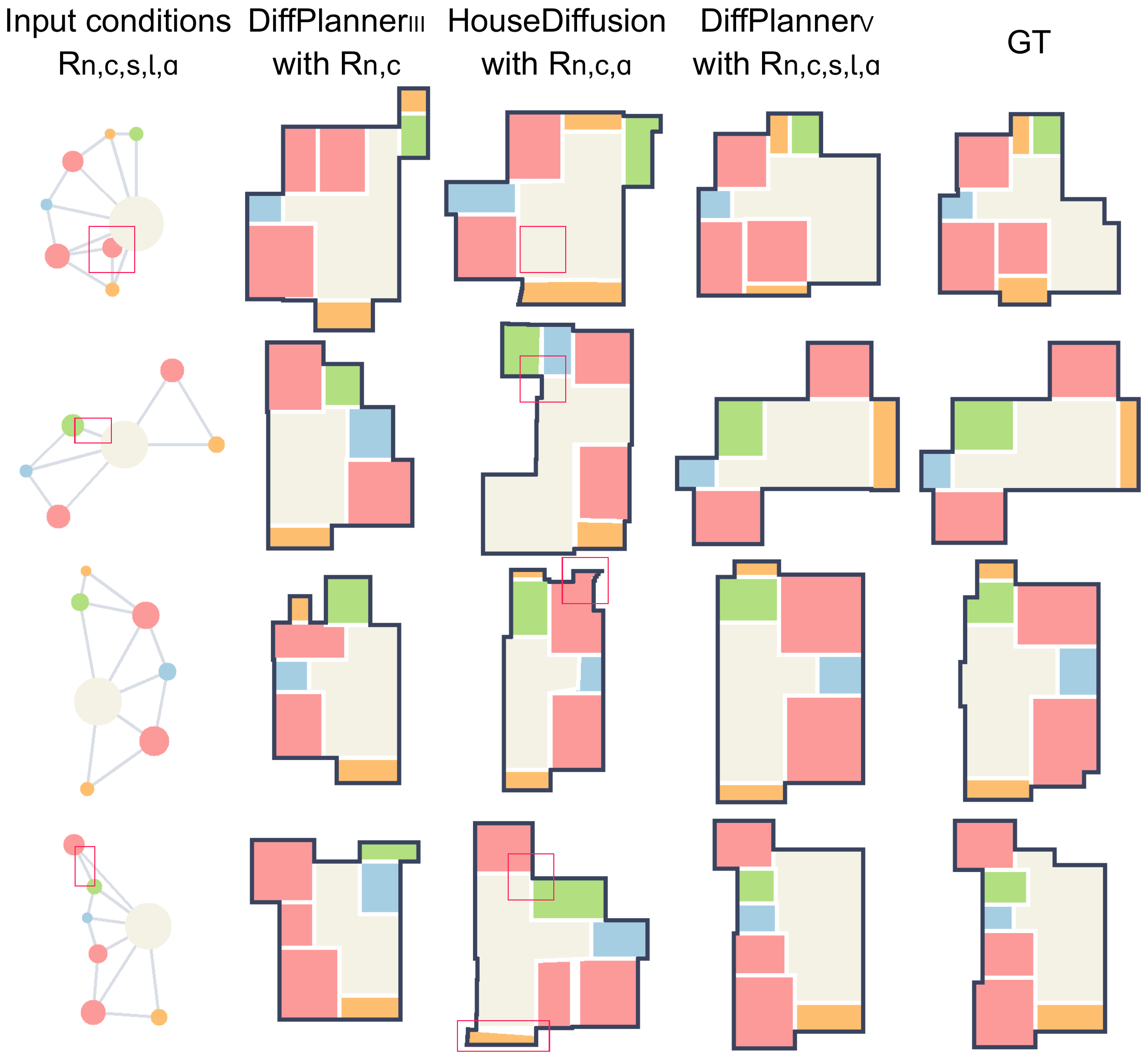}
    \caption{Qualitative comparison of bubble diagram-driven floor plan generation between our method and HouseDiffusion. HouseDiffusion often produces unreasonable results like missing rooms and noisy boundaries. In contrast, our method generates realistic layouts with just the number and categories of rooms ($\text{DiffPlanner}_{\text{III}}$), and closely matches the ground truths (GT) when additional information is provided ($\text{DiffPlanner}_{\text{V}}$).}
    \label{fig:qualitative_fp_wob}
\end{figure}

Our method can also be applied to another prominent task: bubble diagram-driven floor plan generation.
We conducted a qualitative comparison with the state-of-the-art method, HouseDiffusion~\cite{Shabani-23}.
Although HouseDiffusion can predict floor plans based on the number ($R_n$), categories ($R_c$), and adjacencies ($R_a$) of rooms provided in a bubble diagram, it still tends to produce some unreasonable results, such as missing rooms, illogical adjacencies, and noisy boundaries, as shown in Figure~\ref{fig:qualitative_fp_wob}.
In contrast, our $\text{DiffPlanner}_{\text{III}}$ can generate reasonable layout results even when only the number ($R_n$) and categories ($R_c$) of rooms are provided as inputs.
When additional information such as the sizes ($R_s$), locations ($R_l$), and adjacencies ($R_a$) of rooms is added, our $\text{DiffPlanner}_{\text{V}}$ produces results that are almost identical to the ground truths (GT).
This demonstrates the robustness and accuracy of our approach in generating high-quality and realistic floor plans from bubble diagrams.

\begin{figure}[t]
    \centering
    \includegraphics[width=\linewidth]{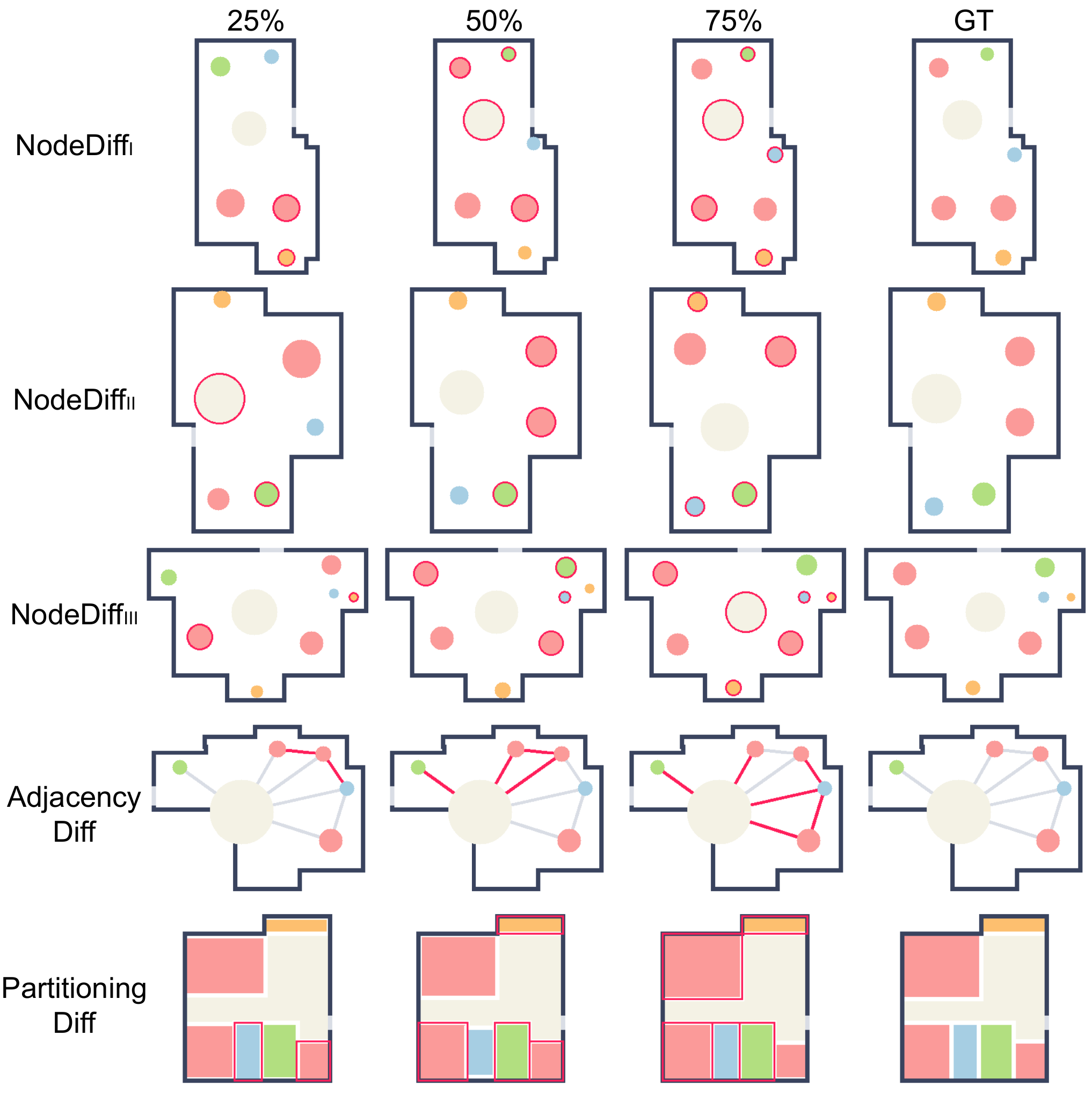}
    \caption{Our three components (NodeDiff, AdjacencyDiff, \& PartitioningDiff) can predict complete layouts from partial inputs with varying proportions ($25\%$, $50\%$, \& $75\%$), enabling detailed user interaction and fine-grained iterative design. Each component produces high-quality results that closely match the ground truths (GT) from the partial inputs. Elements with red borders indicate user-specified partial inputs.}
    \label{fig:qualitative_partial_wob}
\end{figure}

\subsubsection{Iterative design with partial input}

Real-world floor plan layout planning is an iterative design process, making it crucial for a data-driven layout planning tool to support user-driven iterative design.
Unlike previous methods that decompose the generation process into several sub-tasks, allowing users to modify intermediate results for rough control, our goal is to enable iterative design at each sub-task.
This ultimately allows users to have fine-grained control over the entire design process, from initial concept to finalization, for every element in the layout.
The key to achieving this goal is equipping the model with the ability to predict the complete layout from partial user inputs at each sub-task.

Figure~\ref{fig:qualitative_partial_wob} demonstrates the ability of our three components (NodeDiff, AdjacencyDiff, \& PartitioningDiff) to predict complete layouts from partial inputs with varying proportions ($25\%$, $50\%$, \& $75\%$).
The input conditions used by each component are listed in Table~\ref{tab:partial_input}.
Each component performs exceptionally well, predicting high-quality results that closely match the ground truths (GT) from the partial inputs.
This indicates that our approach enables detailed user interaction with the model throughout the design process, supporting fine-grained iterative design.

\begin{figure}[t]
    \centering
    \includegraphics[width=\linewidth]{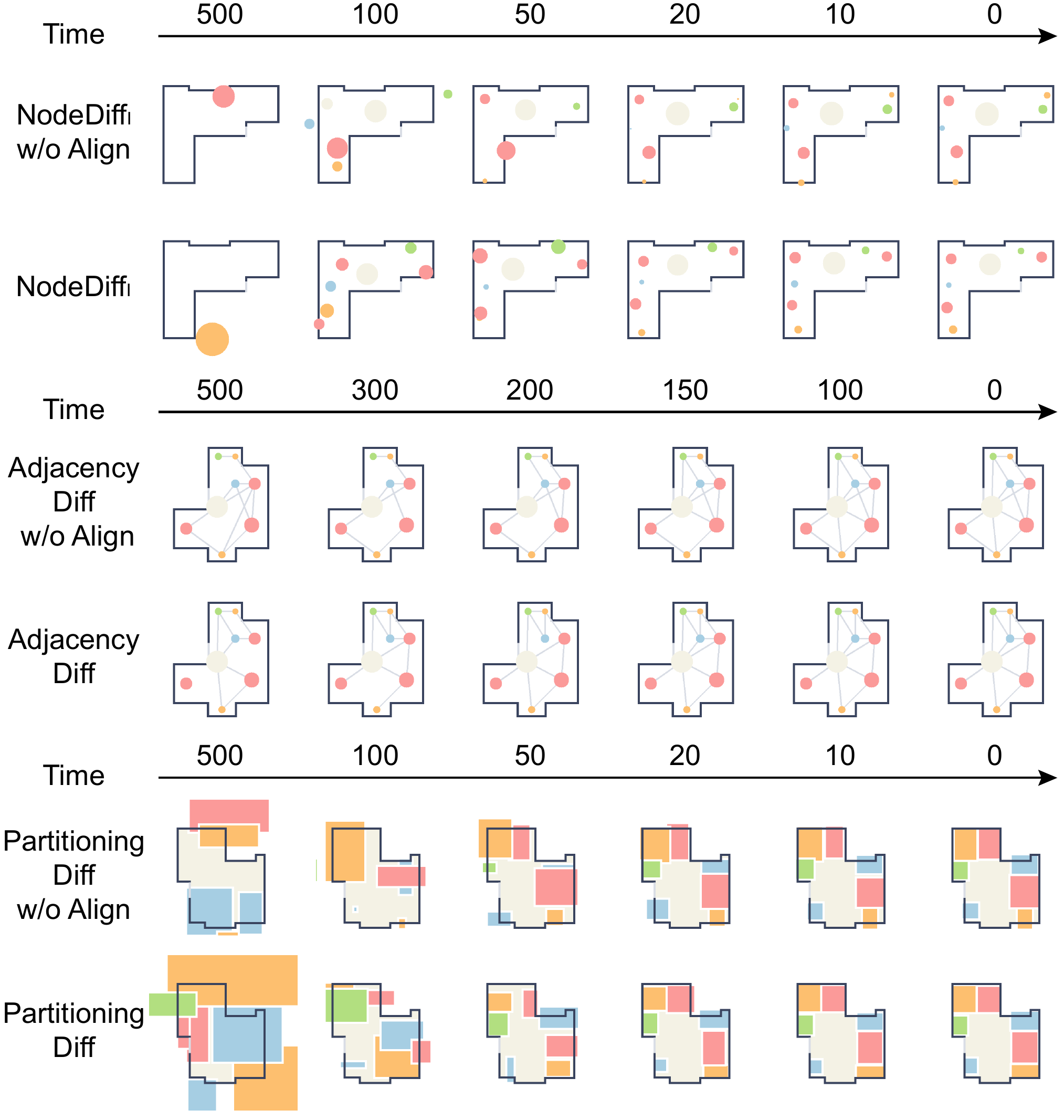}
    \caption{Visual comparisons of the intermediate outputs produced at various diffusion steps by our three core components ($\text{NodeDiff}_{\text{I}}$, AdjacencyDiff, \& PartitioningDiff) and their variants without the alignment mechanism (Align).}
    \label{fig:qualitative_ablation}
\end{figure}

\subsection{Ablation Experiments} \label{subsec:ablation}

We conduct an ablation experiment on the alignment mechanism for the three core components of our method ($\text{NodeDiff}_{\text{I}}$, AdjacencyDiff, \& PartitioningDiff) to demonstrate its effectiveness both qualitatively and quantitatively.

\subsubsection{Qualitative comparison}

Figure~\ref{fig:qualitative_ablation} presents visual comparisons of the intermediate outputs produced at various diffusion steps by our three core components ($\text{NodeDiff}_{\text{I}}$, AdjacencyDiff, \& PartitioningDiff) and their variants without the alignment mechanism ($\text{NodeDiff}_{\text{I}}$ w/o Align, AdjacencyDiff w/o Align, \& PartitioningDiff w/o Align).
The input conditions used by these components and their variants are listed in Table~\ref{tab:ablation_align}.
The comparison results indicate that the alignment mechanism allows our model to more quickly and effectively approximate the ground truth distribution.

For the room node generation task, $\text{NodeDiff}_{\text{I}}$ with the alignment mechanism can already generate a preliminary result at step $50$ that reflects the geometry and topology of the final output.
By step $20$, it essentially produces results consistent with the final output and continues to refine the output in subsequent diffusion steps, finally producing geometrically and topologically coherent room nodes at step $0$.
In contrast, $\text{NodeDiff}_{\text{I}}$ w/o Align only starts to produce results preliminarily consistent with the geometry and topology of the final output at step $20$, achieving a result similar to the final output at step $0$ only by step $10$.
Comparing the results at step $0$, $\text{NodeDiff}_{\text{I}}$ with the alignment mechanism clearly generates room nodes with better geometry and topology, while $\text{NodeDiff}_{\text{I}}$ w/o Align produces room nodes close to or even on the boundary, which is uncommon in the dataset.

Once the room nodes are established, the target distribution of room adjacencies is easily captured by the model.
Therefore, both AdjacencyDiff and AdjacencyDiff w/o Align produce results consistent with the final output at step $0$ by step $100$.
However, compared to AdjacencyDiff w/o Align, AdjacencyDiff with the alignment mechanism is still faster, achieving a result very close to that at step $0$ by step $150$.

For the prediction of room partitioning, PartitioningDiff with the alignment mechanism has essentially moved all room boxes inside the boundary by step $50$.
Although at this time the rooms are not well-aligned and the shapes are somewhat irregular, it further refines the results, eventually producing reasonably distributed room boxes at step $0$.
In contrast, PartitioningDiff w/o Align has still not completely moved all room boxes inside the boundary by step $20$, and by step $0$ it produces some impractical designs, such as the balcony that extends beyond the boundary and does not fully occupy the corners (depicted in yellow located at the bottom right corner), as shown in Figure~\ref{fig:qualitative_ablation}.

\subsubsection{Quantitative comparison}

We further conduct quantitative comparisons of the intermediate outputs produced at various diffusion steps by our three core components ($\text{NodeDiff}_{\text{I}}$, AdjacencyDiff, \& PartitioningDiff) and their variants without the alignment mechanism ($\text{NodeDiff}_{\text{I}}$ w/o Align, AdjacencyDiff w/o Align, \& PartitioningDiff w/o Align).
We use the Fréchet Inception Distance (FID) as the evaluation metric because it assesses the similarity in overall distribution between the generated results and real data.

As shown in Table~\ref{tab:ablation_align}, the FID scores for all three components decreased at various steps when the alignment mechanism was included, indicating an improvement in performance.
The results clearly demonstrate that the alignment mechanism enables faster and more accurate approximation of the ground truth distribution, and enhances the capability of model to generate layouts that more closely resemble the ground truths.

\begin{table}[t]
    \centering
    \caption{Quantitative comparisons of the intermediate outputs produced at various diffusion steps by our three core components ($\text{NodeDiff}_{\text{I}}$, AdjacencyDiff, \& PartitioningDiff) and their variants without the alignment mechanism (Align).}
    \label{tab:ablation_align}
    \resizebox{\columnwidth}{!}{\begin{tabular}{|l|ccc|cccc|}
        \hline
        & \multicolumn{3}{c|}{Input Condition}  & \multicolumn{4}{c|}{FID Score at Time t} \\
        & $B$ & $R_{n,c,s,l}$  & $R_a$ & t=50 & t=20 & t=10 & t=0 \\ \hline
        $\text{NodeDiff}_{\text{I}}$ w/o Align & \cmark & \xmark & \xmark & 27.22  & 6.27 & 3.76 & 3.38   \\
        $\text{NodeDiff}_{\text{I}}$ & \cmark & \xmark & \xmark & \textbf{25.63} & \textbf{5.11} & \textbf{2.72}  & \textbf{2.31}   \\
        \hline
        & $B$ & $R_{n,c,s,l}$  & $R_a$ & t=200 & t=150 & t=100 & t=0 \\ \hline
        AdjacencyDiff w/o Align & \cmark & \cmark & \xmark & \textbf{0.58} & 0.14 & 0.08 & 0.08   \\
        AdjacencyDiff & \cmark & \cmark & \xmark &  \textbf{0.58} & \textbf{0.12} & \textbf{0.05} & \textbf{0.05} \\
        \hline
        & $B$ & $R_{n,c,s,l}$  & $R_a$ & t=50 & t=20 & t=10 & t=0 \\ \hline
        PartitioningDiff w/o Align  & \cmark & \cmark & \cmark & 16.27 &6.05 &3.03 & 1.71 \\
        PartitioningDiff & \cmark & \cmark & \cmark & \textbf{16.14} &\textbf{5.39} &\textbf{2.22}& \textbf{0.85} \\
        \hline
    \end{tabular}}
\end{table}

\subsection{Perceptual Studies} \label{subsec:user_study}

We conduct the perceptual studies with designers to evaluate whether the bubble diagrams and floor plans generated by our $\text{DiffPlanner}^{b}_{\text{I}}$ are comparable to those created by professional designers.
We invite three professional designers, all experienced in graphic design and floor plan layout planning, to participate in the studies.
Two groups of experiments are set up to evaluate bubble diagrams and floor plans, respectively.
Each group includes $50$ comparison tasks where each pair of bubble diagrams (or floor plans) consists of one from the ground truths (GT) and one generated by our $\text{DiffPlanner}^{b}_{\text{I}}$ with only boundary input.
The participants are not aware of the origin of the bubble diagrams or floor plans (being from GT or $\text{DiffPlanner}^{b}_{\text{I}}$), and are asked to complete the forced-choice comparison task by rating each pair as ``GT better/$\text{DiffPlanner}^{b}_{\text{I}}$ better/equally good".
In total, we obtain $150$ answers ($50$ pairs $\times$ $3$ participants) for each group (bubble diagrams and floor plans).

We further compute Fleiss\textquotesingle~Kappa~\cite{Fleiss-71}, a statistical measure of inter-rater agreement, obtaining $0.2$ for floor plans and $0.1$ for bubble diagrams, indicating slight agreement among raters.
This result is expected, as layout evaluation is inherently subjective, and our studies allow users to make holistic decisions without strict evaluation criteria.
Interestingly, agreement is higher for floor plans than for bubble diagrams, which aligns with expectations.
Designers rely on well-established evaluation criteria for floor plan layouts, such as spatial efficiency, functionality, and circulation, whereas bubble diagrams lack standardized assessment frameworks.

\begin{figure}[t]
    \centering
    \includegraphics[width=\linewidth]{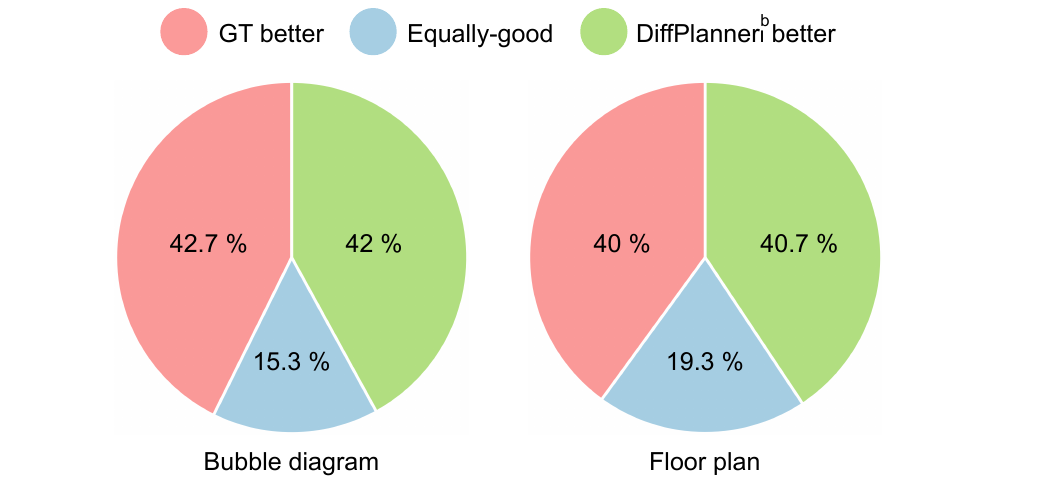}
    \caption{The result of perceptual studies. The results indicate that the bubble diagrams and floor plans generated by our method ($\text{DiffPlanner}^{b}_{\text{I}}$) are nearly indistinguishable from the ground truths (GT).}
    \label{fig:user_study}
\end{figure}

Despite individual user preferences, our perceptual studies are highly successful.
Figure~\ref{fig:user_study} shows the proportion of votes for the results generated by our $\text{DiffPlanner}^{b}_{\text{I}}$ and from the GT in the perceptual studies.
The voting results for both groups indicate that the results generated by our $\text{DiffPlanner}^{b}_{\text{I}}$ are nearly indistinguishable from the GT, demonstrating that our method can produce high-quality, realistic bubble diagrams and floor plans comparable to those designed by professional designers.

\subsection{Discussion} \label{subsec:discussion}

\subsubsection{New data representation}

Previous methods~\cite{Wu-19, Chaillou-20, Hu-20, Wang-21, Sun-22, He-22, Sun-23} often rasterize vector boundaries and entrances into sets of binary raster masks, then utilize image-based networks to extract features.
In contrast, our method feeds the vector boundary and entrance directly into our model, as data tensors and not as images, avoiding the redundant step of rasterizing them to extract their embeddings.
Additionally, we also represent the prediction targets as data tensors, as illustrated in Figure~\ref{fig:overview}(b).
This new representation for both input and output allows our method to operate entirely in the vector space, eliminating the need for rasterization in bubble diagram and floor plan generation tasks.

In our approach, the room categories are mapped to a continuous range of $[-1, 1]$ and represented as a one-dimensional tensor.
This choice is primarily made to maintain a unified input format for diffusion models training and to mitigate the high-dimensional sparsity of one-hot encoding.
Additionally, this normalization stabilizes gradient optimization and facilitates smoother training.
However, we acknowledge that converting categorical variables into continuous values may introduce implicit numerical relationships that do not naturally exist between discrete categories.
Alternative approaches, such as discrete diffusion models~\cite{Austin-21} and hybrid representations, could be explored in future work to better preserve the discrete semantics while maintaining the efficiency of the generative process.

\begin{figure}[t]
    \centering
    \includegraphics[width=\linewidth]{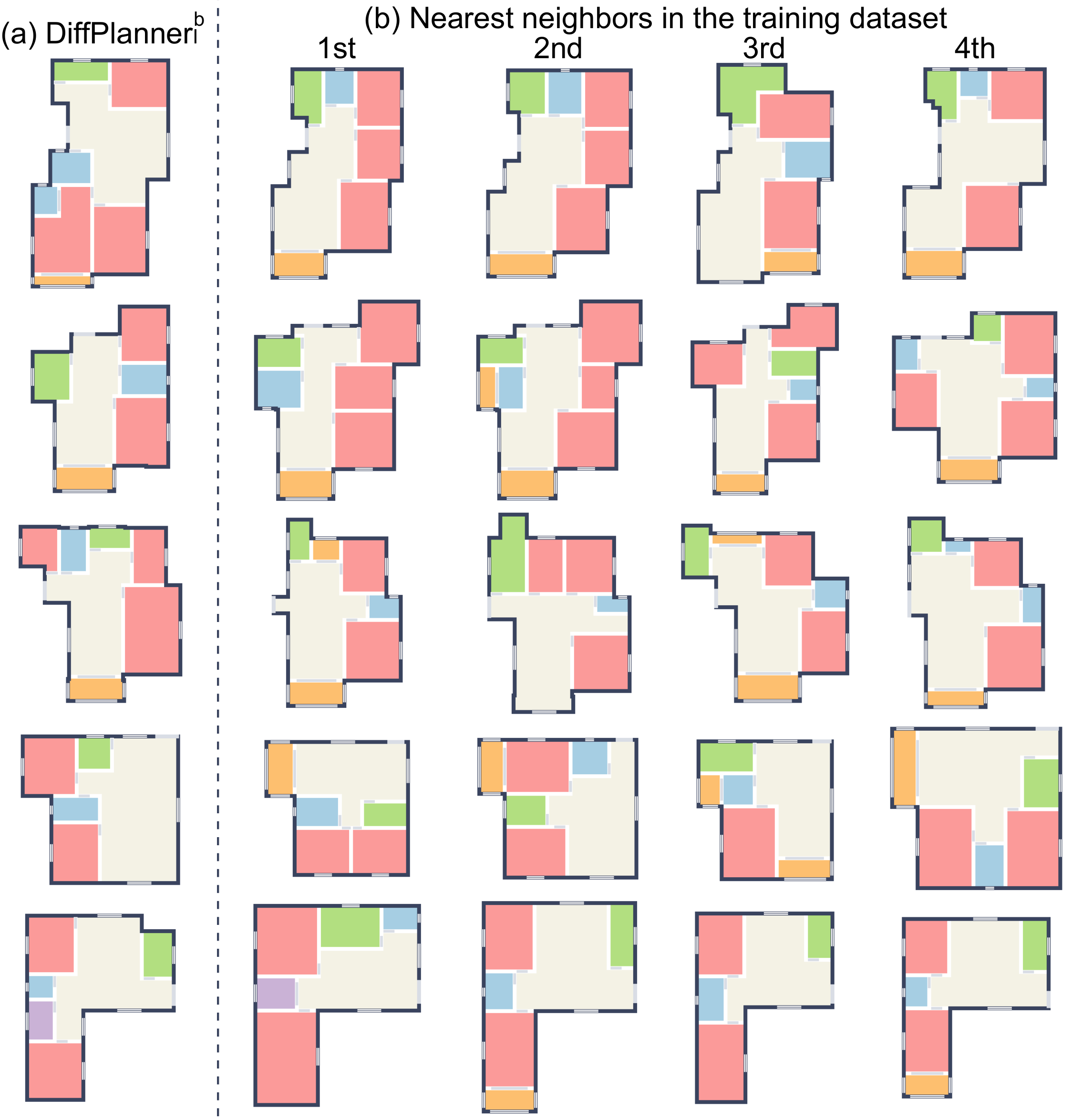}
    \caption{Qualitative comparison between the results generated by our $\text{DiffPlanner}^{b}_{\text{I}}$ based on input samples from the test dataset (a) and their top four retrieved nearest neighbors in the training dataset (b).}
    \label{fig:qualitative_fp_neighbor}
\end{figure}

A natural concern is that data with a tensor-based representation might be too easy to fit, potentially leading to overfitting.
To address this, we conduct both qualitative and quantitative evaluations to demonstrate that our model does not suffer from this risk.
For each sample in the test dataset, we first use $\text{DiffPlanner}^{b}_{\text{I}}$ to generate a corresponding floor plan (shown in Figure~\ref{fig:qualitative_fp_neighbor}(a)) only from the input boundary.
Then, we apply the retrieval algorithm from Graph2Plan~\cite{Hu-20} to find a set of the nearest neighbors from the training dataset based on the similarity of input boundary (shown in Figure~\ref{fig:qualitative_fp_neighbor}(b), with similarity decreasing from left to right).

As illustrated in Figure~\ref{fig:qualitative_fp_neighbor}, the results generated by our $\text{DiffPlanner}^{b}_{\text{I}}$ differ significantly from their nearest neighbors in the training dataset.
This suggests that our method does not suffer from overfitting; rather than simply memorizing the input boundary-to-floor plan mapping and outputting an identical solution to the closest neighbor, our model learns meaningful design patterns from the training data, enabling it to generate reasonable and diverse results.
For example, in the second row of Figure~\ref{fig:qualitative_fp_neighbor}, our $\text{DiffPlanner}^{b}_{\text{I}}$, like its nearest neighbors, places a balcony on the protruding lower-left edge of the input boundary.
However, for other areas, our method explores alternative design choices within the diverse design space instead of replicating the nearest neighbors.
Similarly, in the fifth row of Figure~\ref{fig:qualitative_fp_neighbor}, our method, like its nearest neighbors, places bedrooms in the upper-left and lower-left corners, away from the entrance.
However, the placement of other rooms differs significantly from that of its nearest neighbors in the training dataset, further demonstrating that our model does not overfit but instead generalizes well to new designs.

We also conduct a quantitative comparison of the differences between the result generated by our $\text{DiffPlanner}^{b}_{\text{I}}$ (Our) and its first nearest neighbor (NN) in the training dataset on the areas of six room categories (R) for each sample (s) in the test dataset (S), using the following metric similar to $\text{Coverage}^{GT}_{avg}$ in (\ref{eq:coverage_gt_avg}):

\begin{equation}
    \text{Coverage}^{NN}_{avg} = \left[ \mathbb{E}_{s \in S} \text{IoU}_{s,r}(Our, NN) \right]_{r \in R}.
    \label{eq:coverage_neighbor_avg}
\end{equation}

The $\text{Coverage}^{NN}_{avg}$ ranges from $0$ to $1$, where lower scores indicate less similarity.
As shown in Table~\ref{tab:quantitative_coverage_neighbor}, our $\text{DiffPlanner}^{b}_{\text{I}}$ achieves consistently low $\text{Coverage}^{NN}_{avg}$ scores across all six room categories.
This indicates that the generated results by our method differ significantly from their nearest neighbors in the training dataset, further demonstrating that our method is robust and does not suffer from overfitting.

\begin{table}[t]
    \centering
    \caption{Quantitative comparison of the differences between the result generated by our $\text{DiffPlanner}^{b}_{\text{I}}$ and their first nearest neighbor (NN) in the training dataset on the areas of six room categories, including the living room ($R_{liv}$), bedroom ($R_{bed}$), kitchen ($R_{kit}$), bathroom ($R_{bat}$), balcony ($R_{bal}$), and storage ($R_{sto}$). The $\text{Coverage}^{NN}_{avg}$ ranges from $0$ to $1$, where lower scores indicate less similarity.}
    \label{tab:quantitative_coverage_neighbor}
    \begin{tabular}{|l|cccccc|}
        \hline
        & \multicolumn{6}{c|}{$\text{Coverage}^{NN}_{avg}$}\\
        & $R_{liv}$  & $R_{bed}$  & $R_{kit}$ &$R_{bat}$ & $R_{bal}$ & $R_{sto}$  \\
        \hline
        $\text{DiffPlanner}^{b}_{\text{I}}$ & 0.40 & 0.31 & 0.12  & 0.10  & 0.13 & 0.00 \\
        \hline
    \end{tabular}
\end{table}

\subsubsection{Application scenarios}

Previous methods~\cite{Wu-19, Chaillou-20, Hu-20, Wang-21, He-22, Sun-22} were either specifically designed for boundary-constrained floor plan generation or for bubble diagram-driven floor plan generation.
Among them, WallPlan~\cite{Sun-22} utilized a retrieval approach to enable bubble diagram-driven floor plan generation.
This approach involved adopting a retrieval method to match the input bubble diagram and retrieve the corresponding boundary, followed by a boundary-based layout prediction process.
That is, it is still a boundary-driven floor plan design method.
Different from previous methods, however, we directly treat the vector boundary as a condition.
By simply removing the boundary condition, our method can be easily adapted to generate floor plans without boundary constraints, and the entire prediction process will proceed without any involvement of boundaries.

\subsubsection{Decomposition strategy}

In floor plan generation, previous methods~\cite{Wu-19, Hu-20, Wang-21, He-22, Sun-22} have adopted a decomposition strategy, which breaks down the floor plan generation into several sub-tasks.
This is a universal strategy that has been proven effective in improving the quality of the generated results as well as enhancing user controllability.
In the floor plan domain, the discussion on how to improve the decomposition strategy to enable better user interaction is ongoing.

Our approach differs from previous works in the following ways: We further modified the decomposition strategy for floor plans, starting with room nodes, then room adjacencies, and finally room partitioning.
This allows our model to be used to address two tasks: the data-driven generation of bubble diagrams and floor plans.
This also enables our model to support the maximum user conditions during floor plan design.
Additionally, unlike previous methods that simply decompose the tasks and then allow users to modify the outputs of model before proceeding with subsequent operations to achieve user-controllable generation, our method supports partial input in all three sub-tasks.
This makes our framework fully user-controllable at every step, an achievement not realized by previous methods.

Overall, our method follows the conventional decomposition strategy used in floor plan generation but further updates the strategy to achieve richer user interactions.

\subsubsection{Comparison with HouseDiffusion}

While being closely related and previous state-of-the-art, HouseDiffusion~\cite{Shabani-23} does not take boundary conditions into account, whereas boundaries are a crucial element in floor plan generation~\cite{Wu-19}, which our method accommodates.
Secondly, in the generation of floor plans without boundary constraints, HouseDiffusion only supports generating floor plans from bubble diagrams.
However, bubble diagrams contain a lot of user-driven creative information, and can almost be considered the final stage of user creativity.
For the initial stages of user creativity, the generation of bubble diagrams, HouseDiffusion does not offer support.
Finally, HouseDiffusion also does not support user-controllable generation; it does not allow for partial user inputs, only supporting the generation of floor plans from a complete bubble diagram.

In summary, unlike HouseDiffusion, which specifically supports the transformation from a complete bubble diagram to a floor plan, our goal is to support the interaction with users throughout the entire generative process in all its stages.
We have thus proposed DiffPlanner to achieve our objectives.
Both qualitative and quantitative comparisons demonstrate that DiffPlanner can generate higher-quality results than HouseDiffusion.

\subsubsection{Limitation \& Future work}

In this paper, we represent a room using the coordinates of its top-left and bottom-right corners.
This representation has proven to be efficient in previous methods~\cite{Hu-20, He-22}, enabling the model to quickly capture the target distribution.
However, gaps invariably exist between room boxes, necessitating a post-processing module to align the boxes and eliminate these gaps.

As shown in Table~\ref{tab:quantitative_fp}, compared to Graph2Plan~\cite{Hu-20}, our method has improved fivefold in generating high-quality floor plans without post-processing ($\text{Graph2Plan}_{\text{III}}$ vs. $\text{DiffPlanner}^{b}_{\text{VI}}$).
Nonetheless, it cannot completely eliminate gaps between boxes, as evidenced in Figure~\ref{fig:qualitative_ablation} where some room boxes generated by PartitioningDiff do not align well.
Therefore, we adopt the post-processing method in Graph2Plan~\cite{Hu-20} for floor plan generation task, which effectively eliminates gaps between generated room boxes.
These limitations may be addressed in future work through the introduction of smarter data representations and more advanced generative models.

In our approach, each conditioning setup requires training a separate network to optimize performance for the specific conditioning scheme.
While effective, this may not be the most efficient solution, especially for multiple conditioning setups.
Developing a unified architecture that flexibly integrates arbitrary conditioning remains an interesting direction for future work.

While our study focuses on a constrained task, floor plan generation in residential buildings with limited number of rooms, it remains highly combinatorial, requiring the model to learn meaningful spatial relationships rather than relying solely on retrieval.
Furthermore, our results demonstrate generalization beyond training examples, producing diverse layouts that align with real-world design principles.
Beyond this specific task, our approach has the potential to extend to broader vector-based generative tasks, such as graphic layout design, urban planning, or vector graphics generation, which similarly benefit from direct structure generation without intermediate rasterization.
Future work could explore adaptations of our model to these domains, further testing its applicability.

Regarding practical design utility, while our method aligns with existing architectural heuristics, its real-world relevance depends on its integration into design workflows.
Conducting user studies with professional designers to assess its effectiveness in real applications is an important direction for future research.
Also, while the bubble diagram serves as an intuitive and effective intermediate representation, we acknowledge that other representations may be better suited for certain application scenarios.
Investigating alternative representations for improved model controllability and generation quality remains an interesting avenue for further exploration.
\section{Conclusion}

In this work, we addressed the challenges of boundary-constrained floor plan generation by proposing DiffPlanner, a novel deep learning framework that operates entirely in vector space.
By leveraging the diffusion models and a Transformer-based noise predictor, DiffPlanner effectively handles complex vector data and produces high-quality floor plans.
Our approach eliminates the need for redundant rasterization processes, avoiding common issues such as information loss and distortion.
DiffPlanner integrates a conditioning mechanism for controlled floor plan generation and an alignment mechanism during training, aligning the optimization of model with the iterative design processes of designers.
This ensures that the model can produce detailed vector layouts from various user inputs, and supports various stages of the design process, from early-stage rough designs to detailed floor plans.

Our extensive experiments, including quantitative comparisons, qualitative evaluations, ablation experiments, and perceptual studies, demonstrate that DiffPlanner outperforms existing state-of-the-art methods in generating both bubble diagrams and floor plans.
It offers enhanced user controllability, supporting fully automatic, coarsely controllable, as well as finely controllable generation modes, thus providing users with richer interactions and higher-quality results.

Overall, our contributions highlight the importance of directly learning from vector data and the potential of diffusion models in architectural design and floor plan generation.
DiffPlanner sets a new standard for boundary-constrained and boundary-unconstrained floor plan generation tasks, showcasing the effectiveness of a vector-to-vector approach in generating realistic and precise floor plans.

\bibliographystyle{abbrv}
\bibliography{bibliography}

\begin{IEEEbiography}[{\includegraphics[width=1in,height=1.25in,clip,keepaspectratio]{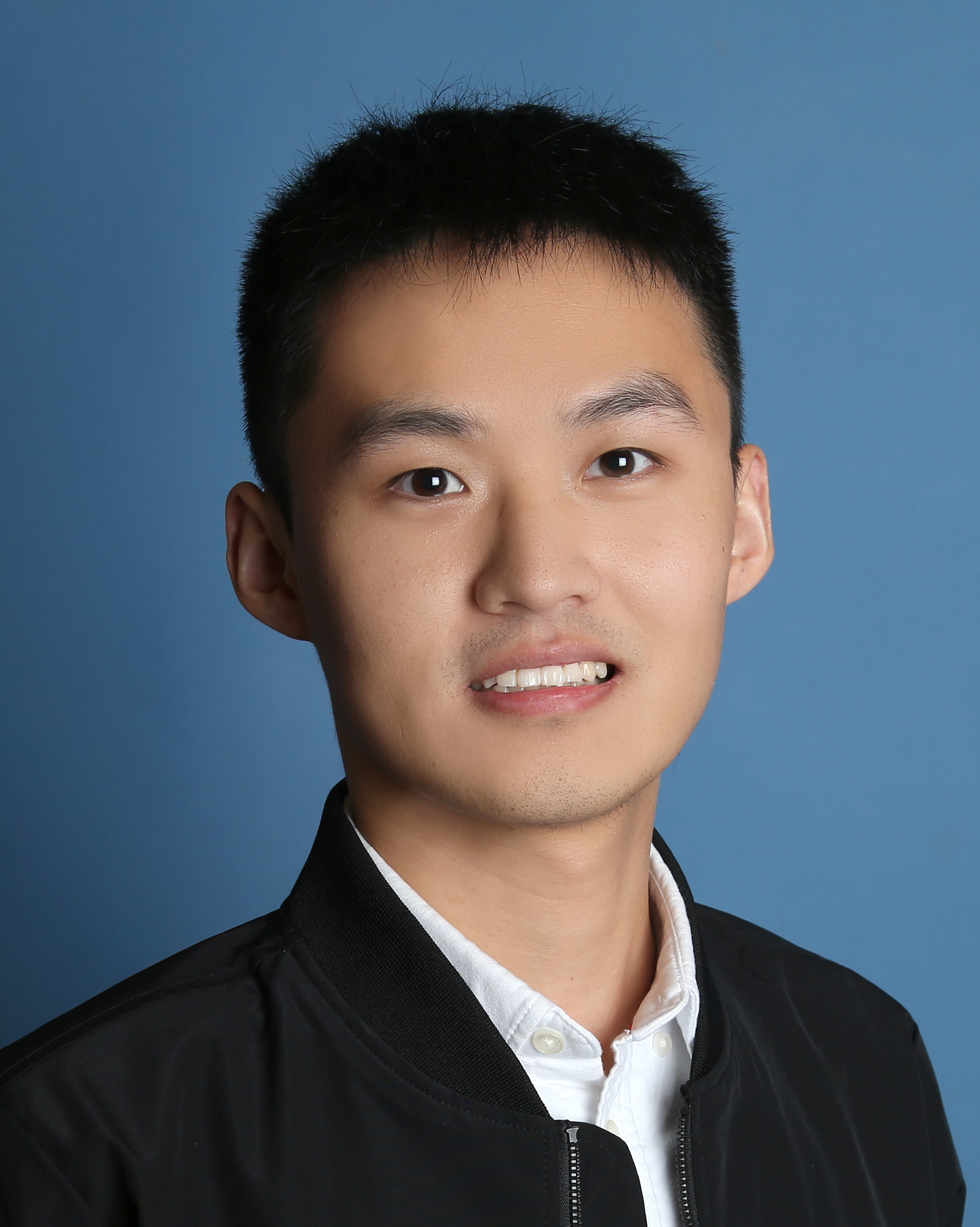}}]
{Shidong Wang} is currently working toward the PhD degree with the Visualization and MultiMedia Lab (VMML), Department of Informatics, University of Zurich.
His research interests include human-AI interaction, deep generative modeling, and computational design.
\end{IEEEbiography}

\begin{IEEEbiography}[{\includegraphics[width=1in,height=1.25in,clip,keepaspectratio]{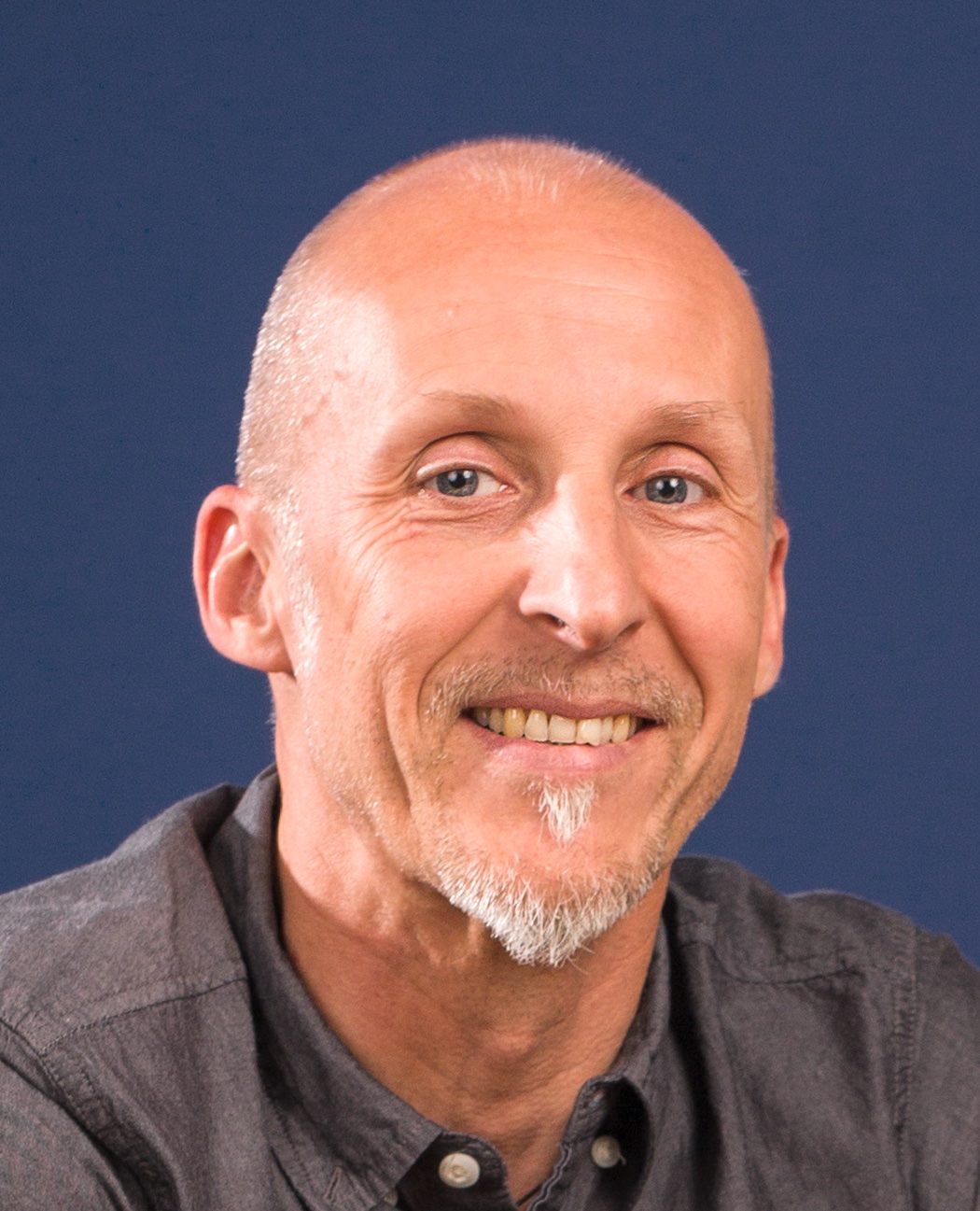}}]
{Prof. Dr. Renato Pajarola}	has been a Professor in computer science at the University of Zurich since 2005, leading the Visualization and MultiMedia Lab (VMML). He has previously been an Assistant Professor at the University of California Irvine and a Postdoc at Georgia Tech. He has received his Dipl. Inf-Ing. ETH and Dr. Sc. techn. degrees in computer science from the Swiss Federal Institute of Technology (ETH) Zurich in 1994 and 1998 respectively. He is a Fellow of the Eurographics Association and a Senior Member of both ACM and IEEE. His research interests include real-time 3D graphics, interactive data visualization, and geometry processing.
\end{IEEEbiography}

\vfill

\end{document}